\documentclass[review,12pt,authoryear]{elsarticle}

\usepackage{amssymb}
\usepackage{amsmath}
\usepackage{graphicx}
\usepackage{epstopdf}
\usepackage{float}
\usepackage{caption}
\usepackage{subcaption}
\usepackage{bm}
\usepackage{bbm}
\usepackage{mathrsfs}
\usepackage{cleveref}
\usepackage{url}
\usepackage{color,soul} %to highlight text
\usepackage{color} %To highlight references in the text
\soulregister\citep7 % To allow citep to be inside of highlighted text
\soulregister\citet7 % Idem with citet
\soulregister\citealp7 % Idem with citealp
\soulregister\ref7 % To allow ref to be inside highlighted text
\soulregister\eqref7 % To allow ref to be inside highlighted text
\soulregister\it7 % To allow it to be inside highlighted text
\font\foef = cmr10 scaled 800

\def\beq{\begin{equation}}
\def\eeq{\end{equation}}
\def\bfig{\begin{figure}}
\def\efig{\end{figure}}
\def\bcn{\begin{center}}
\def\ecn{\end{center}}

\def\bfn{\mbox{\boldmath $n$}}

\def\bfsig{\mbox{\boldmath $\sigma$}}
\def\bfome{\mbox{\boldmath $\omega$}}

\def\bfthe{\mbox{\boldmath $\vartheta$}}

\def\bfgam{\mbox{\boldmath $\gamma$}}

\def\bfh{\mbox{\boldmath $\rho$}}
\def\bfu{\mbox{\boldmath $u$}}

\def\bfzero{\mbox{\boldmath $0$}}

\def\T3{\mbox{$\mathbb T$}}
\def\L4{\mbox{$\mathbb L$}}
\def\M4{\mbox{$\mathbb M$}}

\def\bfT{\mbox{\boldmath $T$}}

\def\bft{\mbox{\boldmath $t$}}

\def\bfeps{\mbox{\boldmath $\varepsilon$}}

\def\bfalp{\mbox{\boldmath $\alpha$}}

\def\bfrho{\mbox{\boldmath $\rho$}}

\def\bfvs{\mbox{\boldmath $\varsigma$}}
\def\bfd{\mbox{\boldmath $\zeta$}}

\def\dive{\mbox{div}}
\def\tr{\mbox{tr}}
\def\curl{\mbox{curl}}
\def\sym{\mbox{sym}}
\def\dvt{\mbox{dev}}
\def\skw{\mbox{skw}}

\def\el{\hbox{\foef el}}

\def\disg{\hbox{\foef dis}}

\newcommand \ds \displaystyle
\newcommand {\de}{\mbox{d}}

\journal{International Journal of Solids and Structures}

\makeatletter
\def\@author#1{\g@addto@macro\elsauthors{\normalsize%
    \def\baselinestretch{1}%
    \upshape\authorsep#1\unskip\textsuperscript{%
      \ifx\@fnmark\@empty\else\unskip\sep\@fnmark\let\sep=,\fi
      \ifx\@corref\@empty\else\unskip\sep\@corref\let\sep=,\fi
      }%
    \def\authorsep{\unskip,\space}%
    \global\let\@fnmark\@empty
    \global\let\@corref\@empty  %% Added
    \global\let\sep\@empty}%
    \@eadauthor={#1}
}
\makeatother

\begin{document}

\begin{frontmatter}



%\title{Modeling the bending of thin foils by distortion gradient plasticity}
%\title{Plane strain boundary value problems in distortion gradient plasticity
%with focus on the bending of thin foils}
%\title{A finite element framework for distortion gradient plasticity with focus on bending of thin foils}
\title{A finite element framework for distortion gradient plasticity with applications to bending of thin foils}


\author{Emilio Mart\'{\i}nez-Pa\~neda\corref{cor1}\fnref{Uniovi}}
\ead{mail@empaneda.com}

\author{Christian F. Niordson\fnref{DTU}}

\author{Lorenzo Bardella\fnref{UBrescia}}

\address[Uniovi]{Department of Construction and Manufacturing Engineering, University of Oviedo, Gij\'on 33203, Spain}

\address[DTU]{Department of Mechanical Engineering, Solid Mechanics, Technical University of Denmark, DK-2800 Kgs. Lyngby, Denmark}

\address[UBrescia]{Department of Civil, Environmental, Architectural Engineering and Mathematics, University of Brescia, Via Branze, 43, 25123 Brescia, Italy}

\cortext[cor1]{Corresponding author. Tel: +34 985 18 19 67; fax: +34 985 18 24 33.}

\begin{abstract}

A novel general purpose Finite Element framework is presented to study small-scale metal plasticity. A distinct feature of the adopted {\it distortion} gradient plasticity formulation, with respect to {\it strain} gradient plasticity theories, is the constitutive inclusion of the plastic spin, as proposed by Gurtin (2004) through the prescription of a free energy dependent on Nye's dislocation density tensor. The proposed numerical scheme is developed by following and extending the mathematical principles established by Fleck and Willis (2009). The modeling of thin metallic foils under bending reveals a significant influence of the plastic shear strain and spin due to a mechanism associated with the higher-order boundary conditions allowing dislocations to exit the body when they reach the boundary. This mechanism leads to an unexpected mechanical response in terms of bending moment versus curvature, dependent on the foil length, if either viscoplasticity or isotropic hardening are included in the model. In order to study the effect of dissipative higher-order stresses, the mechanical response under non-proportional loading is also investigated.

\end{abstract}


\end{frontmatter}


\section{Introduction}
\label{Introduction}

Experiments have shown that metallic materials display strong size effects at both micron and sub-micron scales \citep{F94,NG98,SE98,Moreau05}. Much research has been devoted to modeling the experimentally observed change in the material response with diminishing size \citep{FH97,Q06,K13} in addition to studies of size effects in void growth \citep{L05,N07}, fiber reinforced materials \citep{B03,N03,LN10}, and fracture problems \citep{MB15,MN16}. Most attempts to model size effects in metals have been based on higher-order continuum modeling, and different theories, both phenomenological \citep{FH01,GU04,G04,GA05} and mechanism-based \citep{G99} have been developed. All these theories aim at predicting size effects in polycrystalline metals in an average sense, without explicitly accounting for the crystal lattice, nor for the behavior of internal grain boundaries.\\

While higher-order energetic and dissipative contributions are a common feature among the majority of the most advanced phenomenological strain gradient plasticity (SGP) theories (see, e.g., \citealp{GU04,GA05,GA09,FW09b}), the need to constitutively account for the plastic spin, as proposed about ten years ago by \citet{G04}, to properly describe the plastic flow incompatibility and associated dislocation densities, has been mostly neglected in favor of simpler models. However, the use of phenomenological higher-order formulations that involve the whole plastic distortion (here referred to as \textit{Distortion Gradient Plasticity}, DGP) has attracted increasing attention in recent years due to its superior modeling capabilities. The studies of \citet{BG08} and \citet{B09, B10} have shown that, even for small strains, the contribution of the plastic spin plays a fundamental role in order to provide a description closer to the mechanical response prediction of strain gradient \textit{crystal} plasticity. This has been further assessed by \citet{PP16b}, who, by comparing to a reference crystal plasticity solution given by \citet{GN05}, showed that the plastic rotation must be incorporated to capture the essential features of crystal plasticity. Moreover, \citet{PP16b} numerically elucidated that the localization phenomenon that takes place in \citet{B03} composite unit cell benchmark problem can only be reproduced by DGP. \citet{G04} theory has also been employed by Poh and co-workers \citep{P13,PP16} through a novel homogenization formulation to describe the behavior of each grain in a polycrystal where grain boundaries are modeled to describe effects of dislocation blockage or transmittal. \\

%This has been recently supported by 

However, despite the superior modeling capability of DGP with respect to SGP, the literature is scarce on the development of a general purpose finite element (FE) framework for DGP. Particularly, the use of higher-order dissipative terms - associated with strengthening mechanisms - is generally avoided due to the related computational complexities. This is the case of the very recent FE implementation of \citet{PP16b} and the earlier work by \citet{OG08}, who implemented \citet{G04} theory within a Discontinuous Galerkin framework. Energetic and dissipative contributions are both accounted for in the recent \textit{ad hoc} FE formulation for the torsion problem by \citet{BP15}, also showing that, contrary to higher-order SGP theories, \citet{G04} DGP can predict some energetic strengthening even with a quadratic defect energy.\\

In this work, a general purpose FE framework for DGP is developed on the basis of an extension of the minimum principles proposed by \citet{FW09b}. The numerical scheme includes both energetic and dissipative higher-order stresses and the effect of the latter under non-proportional loading is investigated. The novel FE framework is %then
particularized to the plane strain case and
%used to investigate 
applied to the bending of thin foils, of particular interest to the study of size effects in metals 
(see, e.g., \citealp{YVDGG04, YVDG05, EFPG06, I09, EH09, Pol11}) since the experiments of \citet{SE98} (see also \citealp{Moreau05}). Computations reveal a dependence of the results on the foil length if either rate-dependent plasticity or isotropic hardening are included in the model. This is a consequence of the definition of the energetic higher-order contribution as a function of Nye's dislocation density tensor \citep{N53, FH97, AP99}, that is intrinsic to \citet{G04} theory.
This unexpected effect, absent in conventional theories and in many GP theories, is accompanied with the development of plastic shear and plastic spin, which turn out to 
%be relevant for 
influence the overall mechanical response in bending. Such behavior is triggered by the interaction between the conventional and the higher-order boundary conditions, the latter allowing dislocations to exit the foil at the free boundaries. The foil length dependence of the mechanical response is emphasized by the presence of the plastic spin in \citet{G04} DGP, but it also characterizes the \citet{GA05} SGP theory, still involving Nye's tensor restricted to the assumption of irrotational plastic flow (that is, vanishing plastic spin).
Hence, one of the 
%goals 
results of the present investigation 
%is to shed light on 
concerns with the 
%need to perform 
usefulness of
two-dimensional analyses with
appropriate boundary conditions %in order 
to model %the experimental results in 
micro-bending phenomenologically.
%even in the phenomenological context}.

\paragraph{Outline of the paper}

The DGP theory of \citet{G04} is presented in Section 2, together with the novel minimum principles governing it. The FE formulation and its validation are described in Section 3. Results concerning bending of thin foils are presented and discussed in Section 4. Some concluding remarks are offered in Section 5.

\paragraph{Notation}

We use lightface letters for scalars. Bold face is used for first-, second-, and third-order tensors, in most cases respectively represented by small Latin, small Greek, and capital Latin letters. When we make use of indices they refer to a Cartesian coordinate system. The symbol $``\cdot"$ represents the inner product of vectors and tensors (e.g., $a=\boldsymbol{b} \cdot \boldsymbol{u} \equiv b_i u_i$, $b=\boldsymbol{\sigma} \cdot \boldsymbol{\varepsilon} \equiv \sigma_{ij} \varepsilon_{ij}$, $c=\boldsymbol{T} \cdot \boldsymbol{S} \equiv T_{ijk} S_{ijk}$). For any tensor, say $\boldsymbol{\rho}$, the inner product by itself is $|\boldsymbol{\rho}|^2 \equiv \boldsymbol{\rho} \cdot \boldsymbol{\rho}$. The symbol $``\times"$ is adopted for the vector product: $\boldsymbol{t}=\boldsymbol{m} \times \boldsymbol{n} \equiv e_{ijk} m_j n_k=t_i$, with $e_{ijk}$ denoting the alternating symbol (one of the exceptions, as it is a third-order tensor represented by a small Latin letter), and, for $\boldsymbol{\zeta}$ a second-order tensor: $\boldsymbol{\zeta} \times \boldsymbol{n} \equiv e_{jlk} \zeta_{il} n_k$. For the products of tensors of different order the lower-order tensor is on the right and all its indices are saturated, e.g.: for $\boldsymbol{\sigma}$ a second-order tensor and $\boldsymbol{n}$ a vector, $\boldsymbol{t}=\boldsymbol{\sigma} \boldsymbol{n} \equiv \sigma_{ij} n_j = t_i$; for $\boldsymbol{T}$ a third-order tensor and $\boldsymbol{n}$ a vector, $\boldsymbol{T} \boldsymbol{n} \equiv T_{ijk} n_k$; for $\mathbbmss{L}$ a fourth-order tensor and $\boldsymbol{\varepsilon}$ a second-order tensor, $\boldsymbol{\sigma}=\mathbbmss{L}\boldsymbol{\varepsilon} \equiv L_{ijkl} \varepsilon_{kl} = \sigma_{ij}$. Moreover, $(\nabla \boldsymbol{u})_{ij} \equiv \partial u_i / \partial x_j \equiv u_{i,j}$, $(\textnormal{div} \, \boldsymbol{\sigma})_i \equiv \sigma_{ij,j}$, and $\left( \textnormal{curl} \, \boldsymbol{\gamma} \right)_{ij} \equiv e_{jkl} \gamma_{il,k}$ designate, respectively, the gradient of the vector field $\boldsymbol{u}$, the divergence of the second-order tensor $\boldsymbol{\sigma}$, and the curl of the second-order tensor $\boldsymbol{\gamma}$, whereas $(\textnormal{dev} \, \boldsymbol{\varsigma})_{ij} \equiv (\varsigma_{ij} - \delta_{ij} \varsigma_{kk}/3)$ (with $\delta_{ij}$ the Kronecker symbol), $(\textnormal{sym} \, \boldsymbol{\varsigma})_{ij} \equiv (\varsigma_{ij} + \varsigma_{ji}) /2$, and $(\textnormal{skw} \, \boldsymbol{\varsigma})_{ij} \equiv (\varsigma_{ij} - \varsigma_{ji})/2$ denote, respectively, the deviatoric, symmetric, and skew-symmetric parts of the second-order tensor $\boldsymbol{\varsigma}$.

\section{The flow theory of distortion gradient plasticity and the new stationarity principles}
\label{flth}
The theory presented in this section refers to the mechanical response of a body occupying a space region $\Omega$, whose external surface $S$, of outward normal $\bfn$, consists of two couples of {\it complementary} parts: the first couple consists of $S_t$, where the conventional tractions $\bft^0$ are known,
and $S_u$, where the displacement $\bfu^0$ is known, whereas the second couple consists of $S_t^{\disg}$, where {\it dislocations are free to exit the body}, and $S_u^{\disg}$, where {\it dislocations are blocked and may pile-up}: $S = S_t\cup S_u = S_t^{\disg} \cup S_u^{\disg}$.\\

This section is devoted to the presentation of compatibility, balance, and constitutive equations. For their derivation and for more insight on their mechanical meaning, the reader is referred to \citet{G04} and \citet{B10}.
Furthermore, we will also provide two minimum principles extending those formulated by \cite{FW09b} for a higher-order SGP, to \cite{G04} DGP. On the basis of these minimum principles we will develop the new FE framework in section \ref{FE framework}.

\subsection{Kinematic and static field equations}
\subsubsection{Compatibility equations}

In the small strains and rotations regime, the plastic distortion $\bfgam$, that is the plastic part of the displacement gradient, is related to the displacement $\bfu$ by
\beq
\nabla \bfu = (\nabla \bfu)_{\el}+\bfgam \ \ \ \hbox{in}\ \ \ \Omega
%\nonumber
\eeq
in which $(\nabla \bfu)_{\el}$ is the elastic part of the displacement gradient.
The displacement field $\bfu$ is assumed to be sufficiently smooth, such that $\curl{\nabla \bfu} = \bfzero$ in $\Omega$, and the plastic deformation is assumed to be isochoric, so that $\tr \bfgam =0$.
The total strain, Nye's dislocation density tensor \citep{N53, FH97, AP99},
the plastic strain, and the plastic spin are, respectively, defined as:
\beq
\bfeps = \sym \nabla \bfu\ ,
\ \ \ \ \ \
\bfalp = \curl{\bfgam}\ ,
\ \ \ \ \ \
\bfeps^p = \sym{\bfgam}\ ,
\ \ \ \ \ \
\bfthe^p = \skw{\bfgam}
\ \ \ \ \ \
\ \ \ \hbox{in}\ \ \ \Omega
%\nonumber
\label{kinrel}
\eeq

\subsubsection{Balance equations}

For the whole body free from standard body forces, the conventional balance equation reads
\beq
\dive \bfsig = \bfzero\ \ \ \hbox{in}\ \ \ \Omega
\label{stress_field_1}
\eeq
with $\bfsig$ denoting the standard symmetric Cauchy stress.
\par
The higher-order balance equations
can be conveniently written
into their
symmetric and skew-symmetric parts:
\beq
\bfh-\dvt \bfsig-\dive{\bfT^{(\varepsilon)}}+\sym [\dvt(\curl{\bfd})] = \bfzero
\ \ \ \ \hbox{in}\ \ \Omega
\label{ho1_iso}
\eeq
\beq
\bfome+\skw (\curl{\bfd}) = \bfzero
\ \ \ \ \hbox{in}\ \ \Omega
\label{ho2_iso}
\eeq
in which
$\bfh$, $\bfome$, and $\bfT^{(\varepsilon)}$ are the dissipative stresses constitutively conjugate
to the plastic strain rate $\dot\bfeps^p$, the plastic spin rate $\dot\bfthe^p$, and the
gradient of the plastic strain rate $\nabla\dot\bfeps^p$, respectively, whereas
$\bfd$ is the energetic stress (called %the
 defect stress) constitutively conjugate to Nye's tensor $\bfalp$.
\par
Note that $\bfh$ and $\bfome$ can be added to obtain a dissipative stress, $\bfvs$, conjugate
to the plastic distortion rate $\dot\bfgam$:
\beq
\bfvs = \bfh + \bfome
\ \ \ \ \ \hbox{such that} \ \ \ \ \
\bfh = \sym \bfvs
\ ,\ \ \
\bfome = \skw \bfvs
\ ,\ \ \
\tr \bfvs = 0
%\nonumber
\eeq

\subsection{Boundary conditions}

\subsubsection{Kinematic boundary conditions}
\label{kbcs}
The conventional kinematic boundary conditions are:
\beq
\dot \bfu =\dot \bfu^0\ \ \hbox{on}\ S_u
%\nonumber
\label{u}
\eeq
whereas we adopt homogeneous higher-order kinematic (essential) boundary conditions, which are called microhard boundary conditions as they describe dislocations piling up at a boundary. If the complete DGP theory - including
the third-order dissipative stress $\bfT^{(\varepsilon)}$ - is considered,
the microhard boundary conditions read:
\beq
\dot \bfeps^p=\bfzero\ \ \hbox{and} \ \
\dot \bfthe^p\times \bfn =\bfzero
\ \ \hbox{on}\ S_u^{\disg}
\label{mh1}
\eeq
Otherwise, in the simpler DGP theory neglecting $\bfT^{(\varepsilon)}$,
%only considering, as higher-order stress, the energetic defect stress work conjugate to Nye's tensor,
the microhard boundary conditions read:
\beq
\dot \bfgam\times \bfn =\bfzero
\ \ \hbox{on}\ S_u^{\disg}
\label{mh2}
\eeq

\subsubsection{Static boundary conditions}
The conventional static boundary conditions are:
\beq
\bfsig \bfn=\bft^0 \ \ \hbox{on}\ \ S_t
\label{st_bc}
\eeq
whereas we adopt homogeneous higher-order static (natural) boundary conditions,
which are called microfree boundary conditions as they describe dislocations free to exit the body:
\beq
\bfT^{(\varepsilon)}\bfn+\sym [ \dvt(\bfd \times \bfn)] = \bfzero
\ \ \ \ \hbox{on}\ \ S_t^{\disg}
\label{hobc1}
\eeq
\beq
\skw (\bfd \times \bfn)  = \bfzero
\ \ \ \ \hbox{on}\ \ S_t^{\disg}
\label{hobc2}
\eeq

\subsection{Stationarity principles}
\label{MPs}

In the literature, one of the most common ways to obtain a weak form of the balance equations,
useful for the numerical implementation, is based on the Principle of Virtual Work (see, e.g., \citealp{FH01, GU04, G04}). 
Here, inspired by the work of \citet{FW09a,FW09b},
we instead provide two stationarity principles, leading to the foregoing balance equations, which result in {\it minimum} principles after appropriate constitutive choices are made.
For a given Cauchy stress, the higher-order balance equations \eqref{ho1_iso} and \eqref{ho2_iso} and homogeneous boundary conditions %\eqref{hobc1} and \eqref{hobc2} 
are satisfied by any suitably smooth field $\dot\bfgam$ such that the following functional attains stationarity
\begin{eqnarray}\label{eq:MPIstat}
{\cal H}_1(\dot\bfgam) &=& \int_\Omega \left[ \bfrho\cdot\dot\bfeps^p + \bfome\cdot\dot\bfthe^p
+\bfT^{(\varepsilon)}\cdot\nabla\dot\bfeps^p+
\boldsymbol{\zeta} \cdot \dot{\boldsymbol{\alpha}} - \boldsymbol{\sigma} \cdot \dot{\boldsymbol{\varepsilon}}^p \right]\de V %\\
%\nonumber&&- \int_{S_T}\left[\bfT^{(\varepsilon)}\bfn+\sym [ \dvt(\bfd \times \bfn)]+\skw (\bfd \times \bfn)\right]\cdot\dot\bfgam \de A
\end{eqnarray}
subject to the kinematic relations \eqref{kinrel}.

For a given plastic strain rate, the conventional balance equation \eqref{stress_field_1} and static boundary condition \eqref{st_bc} are satisfied by any kinematically admissible field $\dot\bfu$ that minimizes the following functional:
\begin{equation}\label{eq:MPII}
{\cal J}(\dot{\boldsymbol{u}}) = {1\over 2}\int_\Omega
\mathbbmss{L}\left( \sym\nabla\dot\bfu-\dot{\boldsymbol{\varepsilon}}^{p} \right)
\cdot  \left( \sym\nabla\dot\bfu-\dot{\boldsymbol{\varepsilon}}^{p} \right) \de V - \int_{S_t} \dot{\bft}^0 \cdot \dot{\boldsymbol{u}}\ \de A
\end{equation}

\noindent Here $\mathbbmss{L}$ is the elastic stiffness, relating the elastic strain to the Cauchy stress, $\boldsymbol{\sigma}= \mathbbmss{L} (\boldsymbol{\varepsilon}-\boldsymbol{\varepsilon}^p)$. 

\subsection{Constitutive laws for the energetic terms (recoverable stresses)}

In order to account for the influence of geometrically necessary dislocations (GNDs, see, e.g., \citealp{Ash70, F94, FH97}), the free energy is chosen by \cite{G04} to depend on both the elastic strain, $\boldsymbol{\varepsilon}-\boldsymbol{\varepsilon}^{p}$, and Nye's tensor $\boldsymbol{\alpha}$:
\begin{equation}
\Psi = \frac{1}{2} \mathbbmss{L}  \left( \boldsymbol{\varepsilon}-\boldsymbol{\varepsilon}^{p} \right) \cdot  \left( \boldsymbol{\varepsilon}-\boldsymbol{\varepsilon}^{p} \right) + \mathscr{D} (\boldsymbol{\alpha})
\end{equation}
in which $\mathscr{D} (\bfalp)$ is the so-called \textit{defect energy}, accounting for the plastic distortion incompatibility.
The recoverable mechanisms associated with development of GNDs are incorporated in the current higher-order theory by assuming the following quadratic defect energy:
\begin{equation}
\mathscr{D}(\boldsymbol{\alpha})=\frac{1}{2} \mu \ell^2 \boldsymbol{\alpha}\cdot \boldsymbol{\alpha}
\label{quad_de}
\end{equation}
in which $\mu$ is the shear modulus and $\ell$ is an energetic length scale.
Hence, the defect stress reads:
\begin{equation}
\boldsymbol{\zeta}= \frac{\partial \mathscr{D}(\boldsymbol{\alpha})}{\partial \boldsymbol{\alpha}}=\mu \ell^2 \boldsymbol{\alpha}
\end{equation}
It has been recently shown by \cite{BP15} that it may be convenient to express the defect energy in terms of more
invariants of $\bfalp$, as originally envisaged by \cite{G04}. It may also be relevant to adopt a less-than-quadratic forms of the defect energy (e.g., \citealp{OO07,G10,B10,FG13,BP15}), or even non-convex forms (e.g., \citealp{L15} and references therein). However, the quadratic defect energy is perfectly suitable for the scope of the present investigation, that is implementing \cite{G04} DGP theory in a general purpose FE framework and bringing new features of its
predictive capabilities to attention by analyzing the bending of thin foils.
We leave for further investigations the analysis of other forms of the defect energy.

\subsection{Constitutive laws for the dissipative terms (unrecoverable stresses)}

The unrecoverable stresses are prescribed on the form:
\begin{equation}
\boldsymbol{\rho}=\frac{2}{3} \frac{\Sigma}{\dot{E}^p} \dot{\boldsymbol{\varepsilon}}^p
\ ,\ \ \ \
\boldsymbol{\omega}=\chi \frac{\Sigma}{\dot{E}^p} \dot{\boldsymbol{\vartheta}}^p
\ ,\ \ \ \
\boldsymbol{T}^{(\varepsilon)}=\frac{2}{3} L^2 \frac{\Sigma}{\dot{E}^p} \nabla \dot{\boldsymbol{\varepsilon}}^p
\label{ConstitutiveTerms}
\end{equation}
where the following phenomenological effective plastic flow rate
\begin{equation}
\dot{E}^p=\sqrt{\frac{2}{3} |\dot{\boldsymbol{\varepsilon}}^p|^2+\chi |\dot{\boldsymbol{\vartheta}}^p|^2 + \frac{2}{3}L^2 |\nabla \dot{\boldsymbol{\varepsilon}}^p|^2 }
\label{effflowrate}
\end{equation}
is work conjugate to the effective flow resistance:
\begin{equation}
\Sigma=\sqrt{\frac{3}{2} |\boldsymbol{\rho}|^2+\frac{1}{\chi} |\boldsymbol{\omega}|^2 + \frac{3}{2L^2} |\boldsymbol{T}^{(\varepsilon)}|^2 }
\end{equation}
such that the 2nd law of thermodynamics is satisfied:
\beq
\bfrho\cdot\dot\bfeps^p + \bfome\cdot\dot\bfthe^p + \bfT^{\varepsilon}\cdot\nabla\dot\bfeps^p
\equiv \Sigma\dot E^p > 0 \ \ \forall \ \ \dot\bfgam \ne \bfzero
\label{sat_2nd}
\eeq
In the constitutive laws above $L$ is a dissipative material length parameter and $\chi$ is the material parameter governing the dissipation due to the plastic spin\footnote{By analyzing the simple shear problem, \citet{B09} has provided an analytical expression suggesting that, in order to represent the mechanical response of a {\it crystal}
subject to multislip, $\chi\in [0,2/3]$. However, values of $\chi$ larger than $2/3$ might help in representing the response of crystals in which the plastic flow has preferential orientations.
}.

The form of the function $\Sigma(\dot{E}^p,E^p)$, whose dependence on $E^p$ 
may describe higher-order isotropic hardening, has to be appropriately chosen
to complete the set of constitutive prescriptions
for the unrecoverable stresses.

With these constitutive equations plastic dissipation may be derived from the dissipation potential
\begin{equation}
\mathscr{V} (\dot{E}^p, E^p ) = \int_0^{\ds \dot{E}^p} \Sigma ( e , E^p ) \de e
\end{equation}
which is assumed to be convex in $\dot{E}^p$.
This is important for the development of a numerical solution procedure, as it makes
the stationarity principle based on functional \eqref{eq:MPIstat} a {\it minimum} principle,
whose functional reads:
\begin{equation}\label{eq:MPI}
{\cal H}(\dot\bfgam) = \int_\Omega \left[ \mathscr{V} (\dot{E}^p, E^p ) +
\boldsymbol{\zeta} \cdot \dot{\boldsymbol{\alpha}} - \boldsymbol{\sigma} \cdot \dot{\boldsymbol{\varepsilon}}^p \right]\de V
\end{equation}
Note that in functional \eqref{eq:MPI} $\dot{E}^p$ is a function of $\dot\bfgam$ through equation \eqref{effflowrate}
and the kinematic relations \eqref{kinrel}.

Minimum principles \eqref{eq:MPII} and \eqref{eq:MPI} extend the analogous principles of \citet{FW09b} to the DGP theory of \cite{G04}.

In this work we choose the following viscoplastic potential  
\begin{equation}
\mathscr{V} (\dot{E}^p, E^p ) = {\sigma_Y ( E^p )\dot{\varepsilon}_0 \over m+1}
\left( \dot{E}^p\over \dot{\varepsilon}_0 \right)^{m+1}
\end{equation}
so that
\begin{equation}
\Sigma ( \dot{E}^p , E^p )=\sigma_Y ( E^p )\left( \dot{E}^p \over \dot{\varepsilon}_0 \right)^m
\label{viscoS}
\end{equation}

\noindent with $m$ denoting the rate sensitivity exponent, $\sigma_Y ( E^p )$ the current flow stress given by the % yet unspecified 
hardening rule, and $\dot{\varepsilon}_0$ the reference strain rate.

\section{Finite element formulation}
\label{FE framework}

The present FE framework is based on the minimum principles \eqref{eq:MPII} and \eqref{eq:MPI}. 
%
%The main advantage of the minimum principles is that the plastic strain rate field is directly obtained based on 
%minimum principle I in the context of dissipative gradient effects. For general finite element implementations based on 
%the principle of virtual work (Fredriksson & Gudmundson, 2005, Borg et al., 2007 Niordson & Legarth, 2009) the time derivative 
%of the plastic strain rate field is solved for. This difference makes the present scheme more robust and extended ranges 
%(toward the time independent limit) of the visco-plastic exponent may be analyzed.
%
General finite element implementations of viscoplastic dissipative strain gradient plasticity based on the principle of virtual work (e.g., \citealp{FG05, BNFT06, NL10}) solve for the time derivative of the plastic rate field. The main advantage of employing the minimum principle adopted in the present paper is that the plastic distortion rate field is directly obtained from \eqref{eq:MPI} in the context of dissipative gradient effects. This makes the present numerical scheme more robust as it allows for larger time increments and it enables convergence for lower values of the rate sensitivity exponent. Largely, time-independent behavior may be obtained for sufficiently small rate sensitivity exponents, circumventing complications in the corresponding time-independent model associated with identifying active plastic zones (by, for instance, using image analysis, as proposed by \citealp{NN14}).
%requiring image analysis to define separately active plastic zones needed in the time-independent limit
%associated with tracking the interaction and evolution of plastic zones needed  
Stationarity of \eqref{eq:MPI} together with the constitutive equations \eqref{ConstitutiveTerms} results in the following equation
\begin{eqnarray}\label{eq:variational}
\nonumber&&\int_\Omega \left[\frac{2}{3} \frac{\Sigma}{\dot{E}^p} \dot{\boldsymbol{\varepsilon}}^p\cdot\delta\dot\bfeps^p + \chi \frac{\Sigma}{\dot{E}^p} \dot{\boldsymbol{\vartheta}}^p\cdot\delta\dot\bfthe^p
+\frac{2}{3} L^2 \frac{\Sigma}{\dot{E}^p} \nabla \dot{\boldsymbol{\varepsilon}}^p\cdot\delta\nabla\dot\bfeps^p\right]\de V=\\
&&\int_\Omega \left[\boldsymbol{\sigma} \cdot \delta\dot{\boldsymbol{\varepsilon}}^p-
\boldsymbol{\zeta} \cdot\delta \dot{\boldsymbol{\alpha}}  \right]\de V %\\
%\nonumber&&- \int_{S_T}\left[\bfT^{(\varepsilon)}\bfn+\sym [ \dvt(\bfd \times \bfn)]+\skw (\bfd \times \bfn)\right]\cdot\dot\bfgam \de A
\end{eqnarray}

Given the recoverable stresses, fulfillment of the above weak form (\ref{eq:variational}) of the higher-order equilibrium equations \eqref{ho1_iso} and \eqref{ho2_iso} directly delivers the plastic distortion rate field, $\dot\bfgam$. Adopting Voigt notation, the following FE interpolation is used:
\begin{equation}
\dot{\boldsymbol{u}}=\sum_{n=1}^{N_{I}} \boldsymbol{N}^{(n)} \dot{a}^{(n)}
\end{equation}

\begin{equation}
\dot{\boldsymbol{\gamma}}=\sum_{n=1}^{N_{II}} \boldsymbol{M}^{(n)} \dot{\gamma}^{(n)}
\label{discgamma}
\end{equation}

\noindent Here $\dot{a}^{(n)}=[\dot{a}_1^{(n)} , \dot{a}_2^{(n)}]^T$ and $\dot{\gamma}^{(n)}=[\dot{\gamma}_{11}^{(n)},\ \dot{\gamma}_{22}^{(n)},\ \dot{\gamma}_{12}^{(n)},\ \dot{\gamma}_{21}^{(n)}]^T$ are nodal degrees of freedom and $N_{I}$ and $N_{II}$ are the number of nodes employed for the displacement and the plastic distortion interpolations, respectively. Quadratic shape functions are used for the displacement field ($N_I=8$)
while linear shape functions are employed for the plastic distortion field ($N_{II}=4$). 

Let us note that the continuity requirements for the shape functions related to the 
unconventional FE degrees of freedom are dictated by the structure of 
the kinematic higher-order boundary conditions,
specified in section \ref{kbcs}.
Hence, it is important to point out that we always consider a non-zero dissipative length scale $L$, 
whereas we set it to a very 
small positive number when we want to suppress the effect of the higher-order dissipation.
Therefore, we refer to the higher-order boundary conditions
\eqref{mh1}, which 
%turn out in the requirement that each plastic strain component be continuous in the whole domain.
imply that each plastic strain component must be continuous in the whole domain.
%and its components can be employed as independent 
%%require continuity of the plastic strain and allow the use of its components as independent 
%nodal degrees of freedom. 
This would not be necessarily the case in the theory not accounting for dissipative higher-order 
stresses ($L=0$ in definition \eqref{effflowrate}), in which the shape functions for the unconventional FE degrees of freedom
should be established on the basis of the structure of the boundary conditions \eqref{mh2}. For what concerns the plastic spin,
in the plane strain framework considered in this work there is one single component, so that conditions \eqref{mh1} 
still imply that this component must be continuous in the whole domain.
Overall, the foregoing discussion implies that the four plastic distortion components, adopted as unconventional 
nodal degrees
of freedom as specified in equation \eqref{discgamma}, should be approximated by continuous shape functions.
\par
For general three-dimensional boundary value problems, a totally similar FE framework, in which the eight plastic distortion components 
are employed as unconventional nodal degrees of freedom and are interpolated by continuous shape functions, can be used
by slightly modifying the DGP theory. One should extend the effective plastic flow rate definition \eqref{effflowrate} by including the gradient of the plastic spin rate, weighed by a new dissipative length
scale, say $L_\vartheta$. Of course, with such an extension the DGP theory would be enriched by a further dissipative third-order stress, having nine components, whose divergence would enter the higher-order balance equation
\eqref{ho2_iso}. In this case, the microhard boundary conditions would read
$\dot \bfgam =\bfzero$ on $S_u^{\disg}$. It is uncertain whether the predictive capability of the DGP modelling would largely benefit from such an extension (as inferred by the preliminary analysis in the appendix of \citealp{B10}), but it would be worth investigating, as it leads to a similarly convenient FE implementation as that studied in the present paper.
\footnote{
On the contrary, the \citet{G04} DGP theory involving, as higher-order contribution,
exclusively the defect energy written in terms of
Nye's dislocation density tensor (i.e., $L\equiv 0$ in the theory presented in section \ref{flth}) may be more suitably
implemented in the so-called curl-conforming N\'ed\'elec finite elements \citep{WW11}.
}
\par
Let us finally recall that the static higher-order boundary conditions, specified by equations \eqref{hobc1} and \eqref{hobc2}, are microfree, so that there is no higher-order tractions vector to impose on the boundary $S_t^{\disg}$, where leaving unconstrained an unconventional (plastic) degree of freedom is related to the freedom left to dislocations to exit the body. Dually, setting to zero a plastic degree of freedom on the boundary $S_u^{\disg}$ may trigger plastic distortion gradients, contributing to the size effect through the stiffening of a boundary layer region.

Upon finite element discretization, the weak form (\ref{eq:variational}) of the equilibrium equations \eqref{ho1_iso} and \eqref{ho2_iso} results in a system which is of homogeneous degree zero in terms of the unknown plastic distortion rate field. Imposing the variational form (\ref{eq:variational}) to hold for any kinematically admissible variation of $\dot\bfgam$
leads to the following system of equations, here written in the iterative form (with $l$ denoting the iteration number)
actually implemented:
%Combining \eqref{eq:MPI} with the \hl{constitutive} relations \eqref{ConstitutiveTerms} and the discretization \eqref{discgamma}, while taking into account that
%the weak form holds 
%for any admissible \hl{variation} of $\dot\bfgam$, leads to the following discretized system of equations:
\begin{align}
& \int_{\Omega} \bigg( \frac{\Sigma}{(\dot{E}^p)_{(l-1)}} \Big[ \frac{2}{3} \left( \textnormal{sym} \, \boldsymbol{M}^{(n)} \right) \cdot \left( \textnormal{sym} \, \boldsymbol{M}^{(m)} \right) + \chi \left( \textnormal{skw} \, \boldsymbol{M}^{(n)} \right) \cdot \left( \textnormal{skw} \, \boldsymbol{M}^{(m)} \right)  \nonumber \\
& + \frac{2}{3} L^2  \left( \textnormal{sym} \, \nabla \boldsymbol{M}^{(n)} \right) \cdot \left( \textnormal{sym} \, \nabla \boldsymbol{M}^{(m)} \right)  \Big] \bigg) \de V \cdot (\dot{\gamma}^{(m)})_{(l)} \nonumber \\
& = \int_{\Omega} \Big( \boldsymbol{\sigma} \cdot \left( \textnormal{sym} \, \boldsymbol{M}^{(n)} \right) -  \boldsymbol{\zeta} \cdot \left( \textnormal{curl} \, \boldsymbol{M}^{(n)} \right) \Big) \de V 
\label{systMinI}
\end{align}

%\hl{where $\Omega_e$ is the region occupied by the finite element %, the subscript $l$ denotes the iteration step, and} 
Here the operators $ \textnormal{sym} \, \boldsymbol{M}^{(n)}$, $ \textnormal{skw} \, \boldsymbol{M}^{(n)}$, $\textnormal{sym} \, \nabla \boldsymbol{M}^{(n)}$, and $\textnormal{curl} \, \boldsymbol{M}^{(n)}$
contain the shape functions which deliver the discretizations of $\dot{\boldsymbol{\varepsilon}}^{p}$, $\dot{\boldsymbol{\vartheta}}^{p}$, $\nabla \dot{\boldsymbol{\varepsilon}}^{p}$, and $\dot{\boldsymbol{\alpha}}$, respectively, from the nodal values of the plastic distortion $\dot{\gamma}^{(n)}$ (see \ref{ShapeFunctions}). 
Following \citet{NN11}, 
the system of equations \eqref{systMinI} is solved iteratively for $\dot{\gamma}^{(m)}$
on the basis of the known energetic stresses $(\boldsymbol{\sigma}, \boldsymbol{\zeta}) $ for the current state, written in terms of the total displacement $\boldsymbol{u}$ and plastic distortion $\boldsymbol{\gamma}$ fields at the beginning of the time increment.
%For the first increment the iterative process is initiated by constant guess for the plastic distortion rate field. In subsequent increments, the plastic distortion field 
At a general time increment, the plastic distortion rate field
from the previous increment is used as a starting guess. Convergence of the iteration is defined when the relative norm of the change in the plastic distortion %field is below a threshold value.}
rate field is below an appropriate threshold value.
Finally, the plastic distortion rate $\dot\bfgam$ is determined from the discretization \eqref{discgamma}.

Subsequently, for a known plastic distortion rate field, the incremental solution for the displacement is determined by finding the minimum of functional \eqref{eq:MPII}. The stationarity ensuing from this second minimum principle corresponds to the conventional virtual work statement and, therefore, its implementation into a FE code is standard. Thus, for the sake of brevity, further details are here omitted. In the present incremental procedure we use a Forward Euler time integration scheme, whereas the above described iterative algorithm is implemented so as to ensure convergence in the computation of the plastic distortion rate field. A time increment sensitivity analysis has been conducted in all computations to ensure that the numerical solution does not drift away from the equilibrium configuration.\\

\subsection{Validation of the FE implementation}

In order to validate the present numerical model, the simple shear of a constrained strip is analyzed so as to compare the results with those obtained by \citet{B10} from the minimization of the Total Complementary Energy functional in the deformation theory context. As in \cite{B10}, 
we consider a long strip of height $H$ free from body forces, with isotropic behavior and sheared between two bodies in which dislocations cannot penetrate. 
Hence, the displacement is fully constrained in the lower strip surface, $u_1 (x_2 = 0) = u_2 (x_2 = 0) =0$, while the upper strip surface is subjected to uniform horizontal displacement $u_1 (x_2 = H) = \Gamma H$ with $u_2 (x_2=H)=0$. Here, $\Gamma$ is referred to as the applied strain, 
whose rate, in the following, is assumed to be equal to the adopted reference strain rate ($\dot{\Gamma}=\dot{\varepsilon}_0$). Since dislocations pile-up when they reach the strip lower and upper surfaces, the plastic distortion must be zero at $x_2=0$ and $x_2=H$. The problem is essentially one-dimensional, so that the strip, unbounded along both the shearing direction $x_1$ and the $x_3$ direction, is modeled using a single column of 80 plane strain quadrilateral elements along the strip height ($H$) with appropriate boundary conditions at the sides of the column ($u_2=\gamma_{11}=\gamma_{22}=0\ \forall x_2$).

In order to compare our results with those of \citet{B10}, the following hardening law is used:
\begin{equation}
\sigma_Y ( E^p )=\sigma_0 \left( \frac{E^p}{\varepsilon_0} \right)^N
\label{Powerlaw}
\end{equation}

\noindent We consider the following material properties: $\mu=26.3$ GPa, $\varepsilon_0=0.02$, $\sigma_0=200$ MPa, and $N=0.2$. 

\begin{figure}[H]
\centering
\includegraphics[scale=0.7]{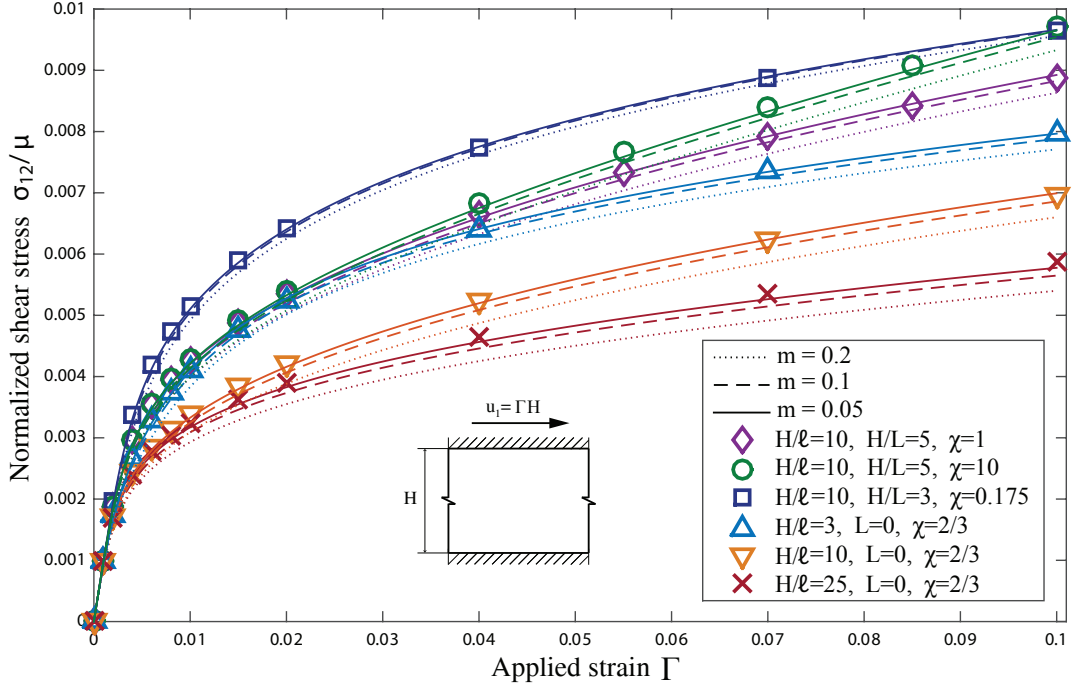}
\caption{Simple shear of a constrained strip. Comparison of the numerical results of the present model (lines) with the predictions of \citet{B10} (symbols) for different values of $H/\ell$, $H/L$, and $\chi$. The case $L=0$ is numerically approximated by setting $L/H=0.01$. Other material parameters are $\sigma_0=200$ MPa, $\varepsilon_0=0.02$, $N=0.2$, $\mu=26.3$ GPa, and $\dot\varepsilon_0$= 0.02 s$^{-1}$.}
\label{fig:Fig1}
\end{figure}

Within the rate-dependent framework adopted, a reference strain rate of $\dot{\varepsilon}_0 = 0.02$ s$^{-1}$ is assumed and the effect of the viscoplastic exponent $m$ is studied in order to approach rate-independent behavior (see equation \eqref{viscoS}). 
Fig. \ref{fig:Fig1} shows the numerical results obtained for different combinations of the material parameter governing the dissipation due to the plastic spin, $\chi$, and the energetic and dissipative length scales, in terms of the ratios $H/\ell$ and $H/L$, respectively. Discrete symbols represent the results obtained by \citet{B10} while solid lines ($m=0.05$), dashed lines ($m=0.1$), and dotted lines ($m=0.2$) show the results of the present FE implementation.\\

As it can be seen in Fig. \ref{fig:Fig1}, the FE framework reproduces the results of \citet{B10} with a very good qualitative and quantitative agreement\footnote{
Note that the results of \citet{B10} are not exact as they were obtained numerically by applying the Rayleigh-Ritz method to the Total Complementary Energy functional. Hence, the present analysis also validates the Rayleigh-Ritz discretization adopted by \citet{B10}.}.

%Since the solution procedure presented in this paper rests on a rate-dependent approach, a very small amount of viscoplastic gradient effects have been included by choosing $L/H=0.01$ for the cases involving exclusively energetic gradient contributions ($L=0$) in \cite{B10}. This facilitates convergence of the proposed FE implementation and has a negligible effect on the mechanical response.\\

\section{Modeling the bending of thin foils}
\label{Results and discussion}

\begin{figure}[H]
        \centering
                \includegraphics[scale=0.7]{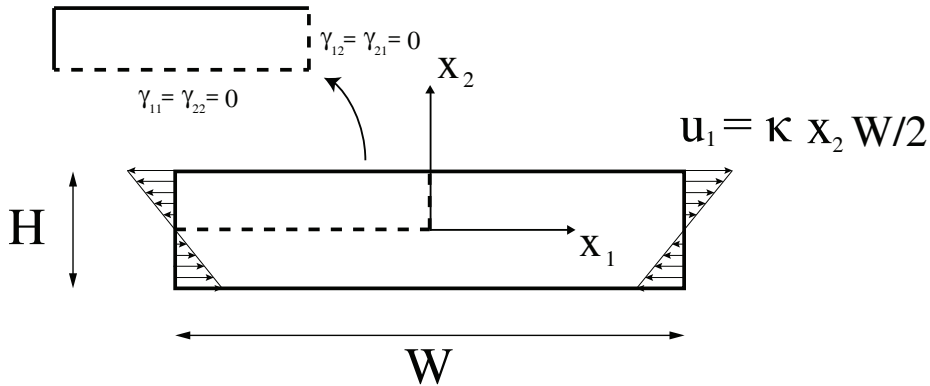}
        \caption{Bending of thin foil: boundary conditions on the undeformed configuration.}\label{fig:Fig2}
\end{figure}

A foil of thickness $H$ and length $W$ subjected to bending is analyzed.
As depicted in Fig. \ref{fig:Fig2}, illustrating the conventional boundary conditions, we impose the longitudinal displacement component at the foil ends:
\begin{equation}
u_1=x_2 x_1 \kappa\ \ \ \hbox{at}\ \ \ x_1\pm W/2
\label{conv_u1}
\end{equation}

\noindent 
whereas the {\it complementary} boundary part is traction-free.
In equation \eqref{conv_u1}, 
$\kappa$ is the curvature that the foil would attain if modeled by a conventional theory,
henceforth referred to as the {\it applied curvature}. 
The higher-order boundary conditions are microfree on the entire
boundary. 
These boundary conditions are adopted for all the monotonic loading analyses.
Note that solving the micro-bending problem as a two-dimensional boundary value problem is 
quite different from what has been done so far in the phenomenological GP literature, 
in which, usually (see, e.g., \citealp{EFPG06, I09, EH09, Pol11}),
the total deformation field is assumed \textit{pointwise} on the basis of the conventional bending theory, thus 
%hindering the specification of the higher-order boundary conditions at the foil ends.
solving for a plastic strain field independent of $x_1$.
Instead, \citet{YVDGG04} and \citet{YVDG05} used %, like us, 
 a two-dimensional plane strain model %ing
to analyze the micro-bending of single crystals by comparing the results of discrete dislocation dynamics 
with those of a backstress-based strain gradient crystal plasticity theory. In both cases, Yefimov et al. 
employ the conventional boundary conditions \eqref{conv_u1} and allow dislocations to exit the foil when
they reach its free boundaries, which corresponds to the microfree boundary condition assumed in this work.
%\hl{A noticeable exception is the two-dimensional plane strain modeling of \citet{YVDGG04} and \citet{YVDG05}, who 
%analyzed the micro-bending of single crystals by comparing the results of discrete dislocation dynamics 
%with those of a backstress-based strain gradient crystal plasticity theory. In both cases, Yefimov et al. 
%use the conventional boundary conditions \eqref{conv_u1} and allow dislocations to exit the foil when
%they reach its free boundaries, that corresponds to the microfree boundary condition assumed in this work.}
\par
As detailed %it will be explained in detail 
in section \ref{sub41}, the structure of the microfree boundary conditions
is the responsible for the need to solve a two-dimensional boundary value problem in order to obtain the 
solution of the micro-bending problem described by the here concerned Nye's tensor-based phenomenological GP.
In particular, we will show that the boundary conditions here adopted lead to a peculiar mechanical response
whose validation would require specific experiments. Moreover, our results imply that modeling actual 
bending experiments
available in literature \citep{SE98,Moreau05} may require two-dimensional analysis and particular attention to 
the boundary conditions to be imposed, the latter being not necessarily those used in this study.%, in which we mostly aim at illustrating both the capability of the new FE framework and the behavior ensuing from microfree boundary conditions in GP based on the plastic distortion incompatibility.

By exploiting symmetry and skew-symmetry conditions of the bending problem, we may impose that:
\begin{equation}
\gamma_{11}=\gamma_{22}=0 \ \ \hbox{at} \ \ x_2=0
\ \ \ \hbox{and} \ \ \
\gamma_{12}=\gamma_{21}=0 \ \ \hbox{at} \ \ x_1=0
\end{equation}
in such a way as to model only one fourth of the foil, as depicted in Fig. \ref{fig:Fig2}.
The vertical displacement of the center node is constrained in order to suppress rigid body motion. 

\subsection{Micro-bending within Nye's tensor-based phenomenological gradient plasticity}
\label{sub41}
%
%The interpretation of the results require the analysis of the governing equations.
 In plane strain problems the sole non-vanishing Nye's tensor components are
 \begin{equation}
 \alpha_{13} = \gamma_{12,1}-\varepsilon_{11,2}^p
 \ , \ \ \
 \alpha_{23} = \varepsilon_{22,1}^p-\gamma_{21,2}
\ , \ \ \
 \alpha_{31} = \varepsilon_{33,2}^p
 \ , \ \ \
 \alpha_{32} = -\varepsilon_{33,1}^p
 \label{nyeplanestrain}
 \end{equation}
At the foil ends the homogeneous microfree boundary conditions \eqref{hobc1} and \eqref{hobc2} provide
\begin{equation}
{2\over 3}L^2{\Sigma\over \dot E^p}\dot\varepsilon_{11,1}^p+
{\mu\ell^2\over 3} (\varepsilon^p_{11,1}+\gamma_{21,2}) = 0\ \ \ \hbox{at}\ \ \ x_1=\pm W/2
\end{equation}
\begin{equation}
{2\over 3}L^2{\Sigma\over \dot E^p}\dot\varepsilon_{12,1}^p+
{\mu\ell^2\over 2} (\gamma_{12,1}-\varepsilon^p_{11,2}) = 0\ \ \ \hbox{at}\ \ \ x_1=\pm W/2
\label{mfreeeps12W}
\end{equation}
\begin{equation}
{2\over 3}L^2{\Sigma\over \dot E^p}\dot\varepsilon_{22,1}^p+
{\mu\ell^2\over 3} (\varepsilon^p_{11,1}+3\varepsilon^p_{22,1}-2\gamma_{21,2}) = 0\ \ \ \hbox{at}\ \ \ x_1=\pm W/2
\end{equation}
\begin{equation}
\gamma_{12,1}-\varepsilon^p_{11,2} = 0\ \ \ \hbox{at}\ \ \ x_1=\pm W/2
\label{mfreethe12W}
\end{equation}
Combination of \eqref{mfreeeps12W} and \eqref{mfreethe12W} leads to
\begin{equation}
\dot\varepsilon_{12,1}^p  = 0\ \ \ \hbox{at}\ \ \ x_1=\pm W/2
\label{comb_ho_bc}
\end{equation}

At the foil top and bottom surfaces the microfree boundary conditions \eqref{hobc1} and \eqref{hobc2} provide
similar relations, among which the most relevant reads:
%\begin{equation}
%{2\over 3}L^2{\Sigma\over \dot E^p}\dot\varepsilon_{22,2}^p+
%{\mu\ell^2\over 3} (\gamma_{12,1}+\varepsilon^p_{22,2}) = 0\ \ \ \hbox{at}\ \ \ x_2=\pm H/2
%\label{mfreeeps12H}
%\end{equation}
%\begin{equation}
%{2\over 3}L^2{\Sigma\over \dot E^p}\dot\varepsilon_{12,2}^p-
%{\mu\ell^2\over 2} (\varepsilon^p_{22,1}-\gamma_{21,2}) = 0\ \ \ \hbox{at}\ \ \ x_2=\pm H/2
%\label{ep122}
%\end{equation}
%
\begin{equation}
{2\over 3}L^2{\Sigma\over \dot E^p}\dot\varepsilon_{11,2}^p-
{\mu\ell^2\over 3} (2\gamma_{12,1}-3\varepsilon^p_{11,2}-\varepsilon^p_{22,2}) = 0\ \ \ \hbox{at}\ \ \ x_2=\pm H/2
\label{mfreeeps11H}
\end{equation}
%\begin{equation}
%\varepsilon^p_{22,1}-\gamma_{21,2} = 0\ \ \ \hbox{at}\ \ \ x_2=\pm H/2
%\label{mfreethe12H}
%\end{equation}
%Combination of \eqref{ep122} and \eqref{mfreethe12H} leads to
%
%\begin{equation}
%\dot\varepsilon_{12,2}^p  = 0\ \ \ \hbox{at}\ \ \ x_2=\pm H/2
%\end{equation}
%
Inspection of the foregoing equations, with particular reference to \eqref{mfreethe12W}, allows us to deduct that, in the plastic regime, at the foil end regions a non-vanishing $\gamma_{12}$ must develop. Furthermore, when $\varepsilon^p_{11,2}$ becomes sufficiently large, an increase of $\gamma_{12,1}$ is expected in order to minimize the defect energy in the end regions (see Nye's tensor component $\alpha_{13}$ in equation \eqref{nyeplanestrain}). {\it This implies that, for a given $H$, the DGP theory here concerned may predict a mechanical response dependent on the foil length $W$.}
Let us notice that the contributions %of the} plastic spin 
of $\vartheta^p_{12}$
and %plastic shear strain 
$\varepsilon^p_{12}$ %\hl{contributing} 
to $\gamma_{12}$ depend on the
chosen material parameters. In particular, $\chi=0$ makes it energetically convenient to develop plastic spin
to minimize the defect energy, while $\chi\to \infty$ leads to the irrotational plastic flow condition of \citet{GA05},
allowing for the development of $\varepsilon^p_{12}$ only.\\

With the aim of gaining insight into the role of both $\varepsilon^p_{12}$ and $\vartheta^p_{12}$ in the bending problem, we have carried out several analyses with the present FE framework. 
%\par
Unless otherwise specified, the ratio 
$W/H=30$
is adopted.
\par
%We will vary the rate sensitivity parameter $m$ in order to discuss how small it should be in order to reproduce nearly rate-independent behavior.
For each case presented different mesh densities were used to ensure achieved convergence. Typically, 20 quadrilateral elements were employed along the thickness and uniform meshes were used, with element aspect ratio equal to 1. 
Both full- and reduced-integration plane strain elements (having, respectively, nine and four Gauss integration points) 
were tested and no shear locking effects were observed. 
For the sake of clarity we will focus our attention to perfectly plastic behavior, that is $N=0$ in equation \eqref{Powerlaw}.\\

Henceforth, we adopt the following material properties: $\mu=26.3$ GPa, Poisson's ratio $\nu=0.3$, initial yield stress $\sigma_0=200$ MPa, and reference strain rate $\dot{\varepsilon}_0=0.02$ s$^{-1}$. Other material parameters will be specified case by case. Unless otherwise stated,
the dissipative and energetic length scales are such that 
\beq
H/L=2.5\ \ \ \hbox{and}\ \ \ H/\ell=5 
\nonumber
\eeq
The specimen is loaded at a rate of curvature $\dot{\kappa} =  \sqrt{3} \dot{\varepsilon}_0 / H$, 
such that, in conventional bending,
the most stretched material points would be loaded at a conventional effective plastic strain rate equal to $\dot{\varepsilon}_0$
when elastic strain increments vanish.

\begin{figure}[H]
\centering
\noindent\makebox[\textwidth]{%
\includegraphics[scale=0.35]{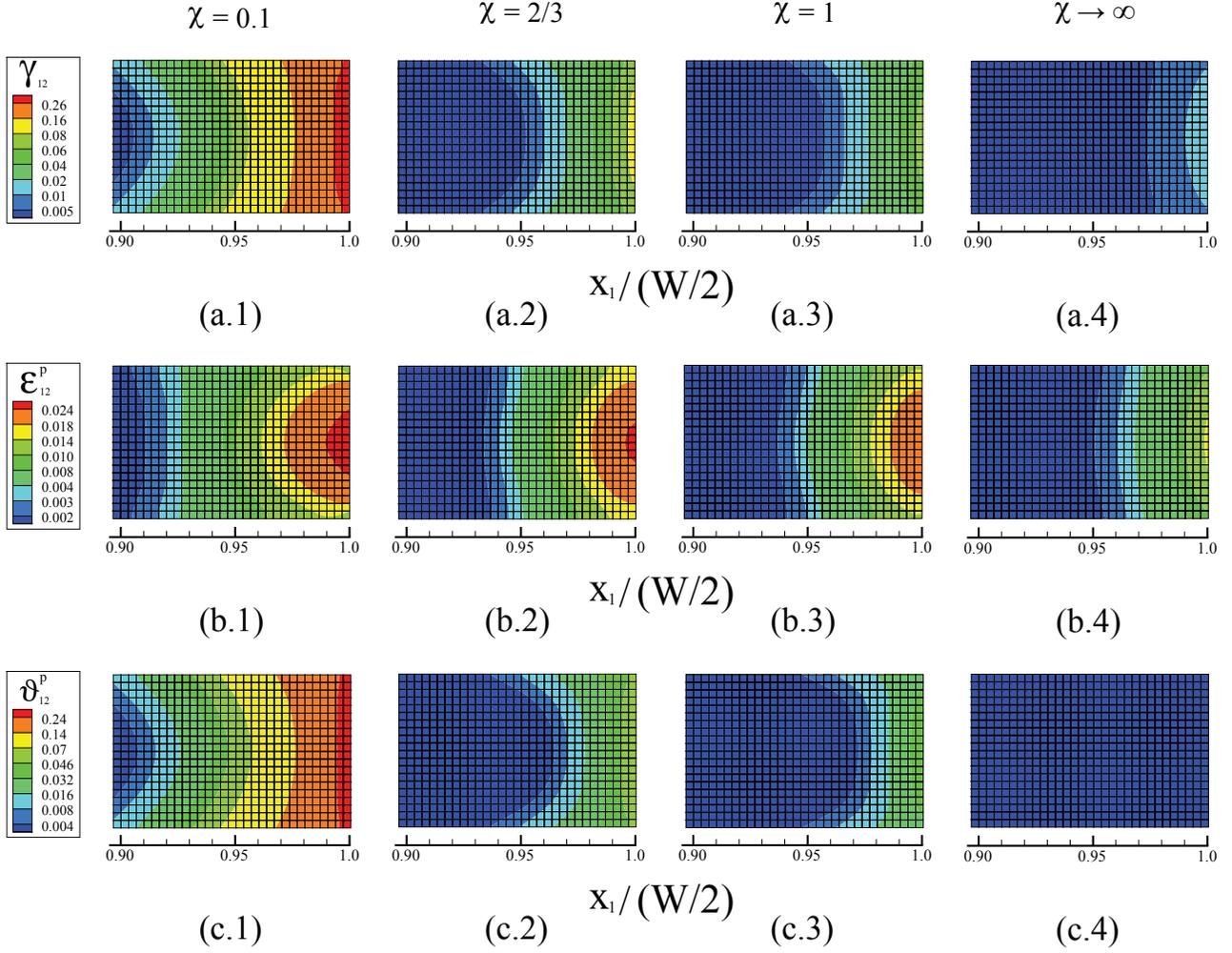}}
\caption{Contours of $\gamma_{12}$ (a), $\varepsilon_{12}^p$ (b), and $\vartheta_{12}^p$ (c) at $ H \kappa / \sqrt{3} =0.05$ for $\chi=0.1$ (1), $2/3$ (2), $1$ (3), 
and $\chi \to \infty$ (4). The rate sensitivity exponent is $m=0.05$.}
\label{fig:Fig3}
\end{figure}

Fig. \ref{fig:Fig3} represents the contours obtained for $\gamma_{12}$, $\varepsilon^p_{12}$, and $\vartheta^p_{12}$ at the applied normalized curvature 
$ H \kappa / \sqrt{3} =0.05$. The influence of different values of $\chi$ is examined by adopting $\chi=0.1$, $\chi=2/3$, $\chi=1$, and $\chi \to \infty$. $\chi=2/3$ is an upper limit estimate to represent crystal multislip \citep{B09} and makes the effective plastic flow rate \eqref{effflowrate} equal to the norm of the plastic distortion in the absence of dissipative higher-order terms. $\chi \to \infty$ reproduces the conditions of \citet{GA05} SGP theory.\\

\begin{figure}[H]
\centering
\noindent\makebox[\textwidth]{%
\includegraphics[scale=0.95]{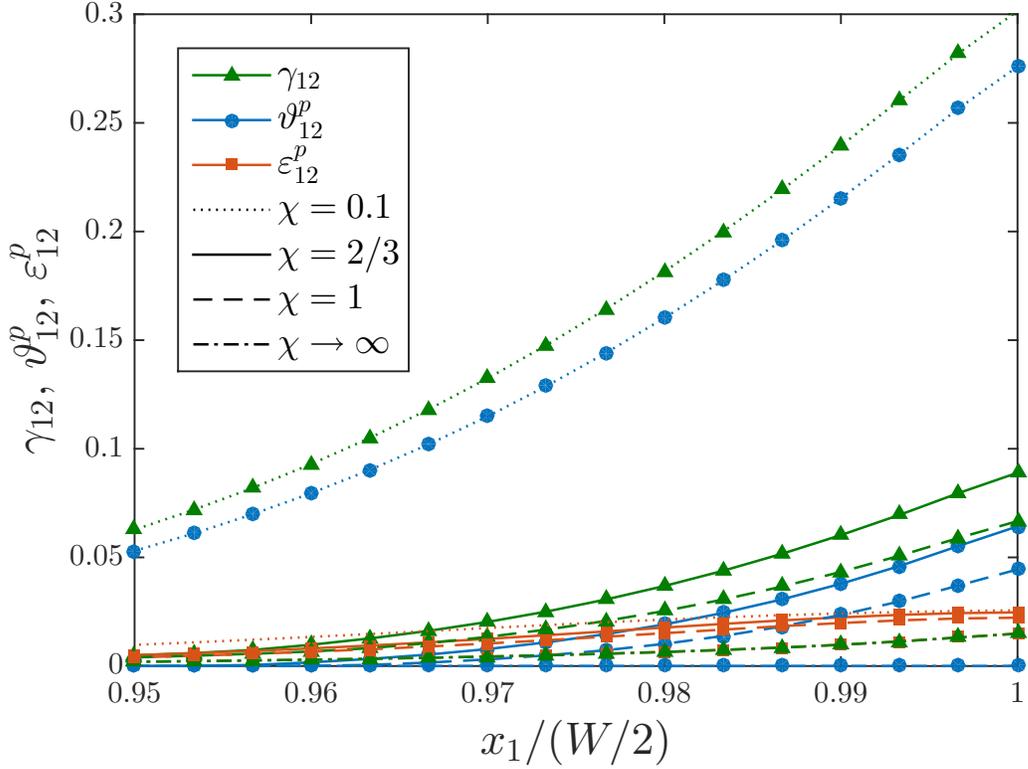}}
\caption{Variation of $\gamma_{12}$, $\varepsilon_{12}^p$, and $\vartheta_{12}^p$ along $x_1$ ($x_2=0$) at  $H \kappa / \sqrt{3}=0.05$ for $\chi=0.1$, $2/3$, 
$1$, and $\chi \to \infty$. The rate sensitivity exponent is $m=0.05$.}
\label{fig:Fig4}
\end{figure}

The results reveal a strong influence of $\gamma_{12}$ (Fig. 3(a.1)-(a.4)), which increases towards the foil end. 
Unexpected within a classical framework, both $\varepsilon_{12}^p$ (Fig. 3(b.1)-(b.3)) and $\vartheta_{12}^p$ (Fig. 3(c.1)-(c.3)) assume relevant values in a significant foil region. Their role is weighed by the value of $\chi$, with $\varepsilon_{12}^p$ increasing notably as $\chi$ decreases. 
The variations of $\gamma_{12}$, $\varepsilon^p_{12}$, and $\vartheta^p_{12}$ can be better appreciated in Fig. \ref{fig:Fig4}, where they are plotted as functions of the foil axis $x_1$.

\begin{figure}[H]
\centering
\noindent\makebox[\textwidth]{%
\includegraphics[scale=0.95]{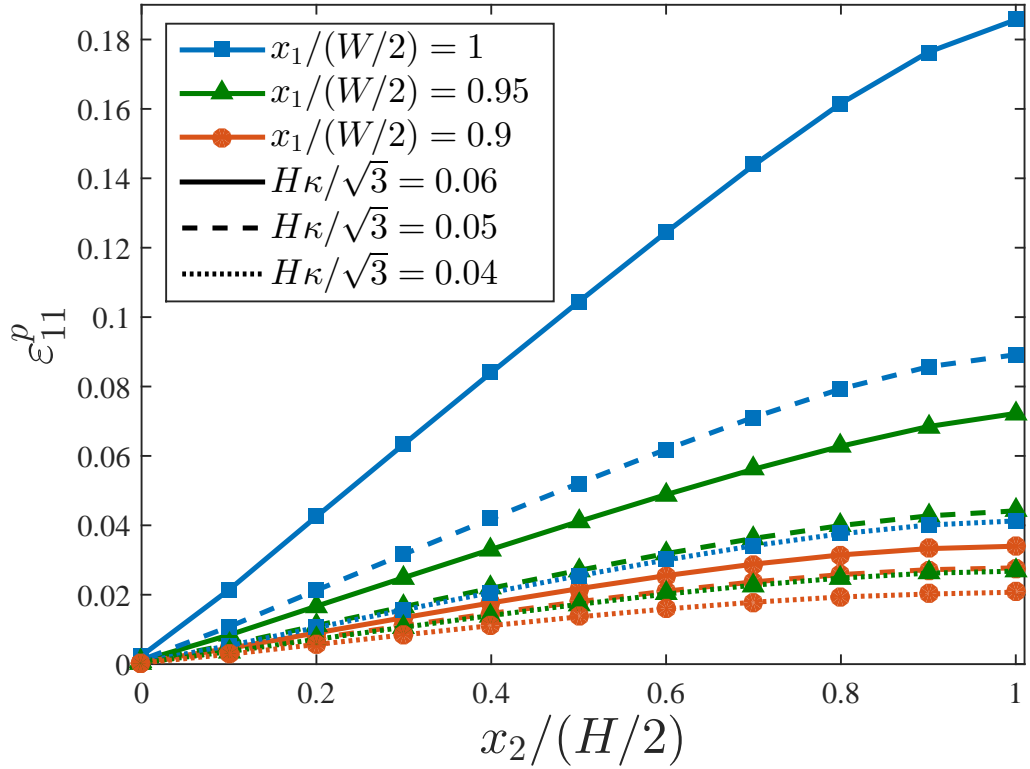}}
\caption{Variation of $\varepsilon_{11}^p$ along $x_2$ in different foil cross-sections at different applied curvature values. The following material properties are adopted: $\chi=2/3$ and $m=0.05$.}
\label{fig:Fig5}
\end{figure}

As it can be seen in Fig. \ref{fig:Fig4}, in all cases $\gamma_{12}$, $\varepsilon_{12}^p$, and $\vartheta_{12}^p$ are monotonic functions of $x_1$, reaching the maximum at the foil end. 
Again, we observe that the contribution of $\vartheta_{12}^p$ to $\gamma_{12}$ becomes dominant
as $\chi$ decreases towards zero. Regarding $\varepsilon_{12}^p(x_1)$ one must note that %, although it increases along $x_1$, 
there is a notable decrease in its slope for $x_1\to W/2$. %in the vicinity of the foil end. 
This is a consequence of the homogeneous microfree boundary conditions, requiring $\varepsilon_{12,1}=0$ at $x_1=W/2$ (see eq. \eqref{comb_ho_bc}). The peculiar development of $\gamma_{12}$ at the foil end is due to the need of accommodating $\varepsilon_{11,2}^p$, as expressed by eq. \eqref{mfreethe12W}. 
In this region, when $\kappa$ is large enough $\varepsilon_{11}^p$ strongly varies with $x_1$, as shown in Fig. \ref{fig:Fig5}. Here, contrary to conventional plasticity, $|\varepsilon_{11,2}^p|$ increases with $|x_1|$.
%As the applied curvature increases in the plastic regime, $\varepsilon_{11}^p$ starts to vary significantly along the foil axis $x_1$, as shown by Fig. \ref{fig:Fig5}.\\ 

%Although $\varepsilon_{11}^p$ varies through the thickness of the foil in all sections, Fig. \ref{fig:Fig5} reveals noticeable differences among sections in the foil end region: contrary to conventional plasticity, 
%$|\varepsilon_{11,2}^p|$ significantly increases with $|x_2|$ close to the foil ends.
%Fig. \ref{fig:Fig5} also represents the expected increase of $|\varepsilon_{11,2}^p|$ with the applied curvature.

\begin{figure}[H]
\centering
\noindent\makebox[\textwidth]{%
\includegraphics[scale=0.95]{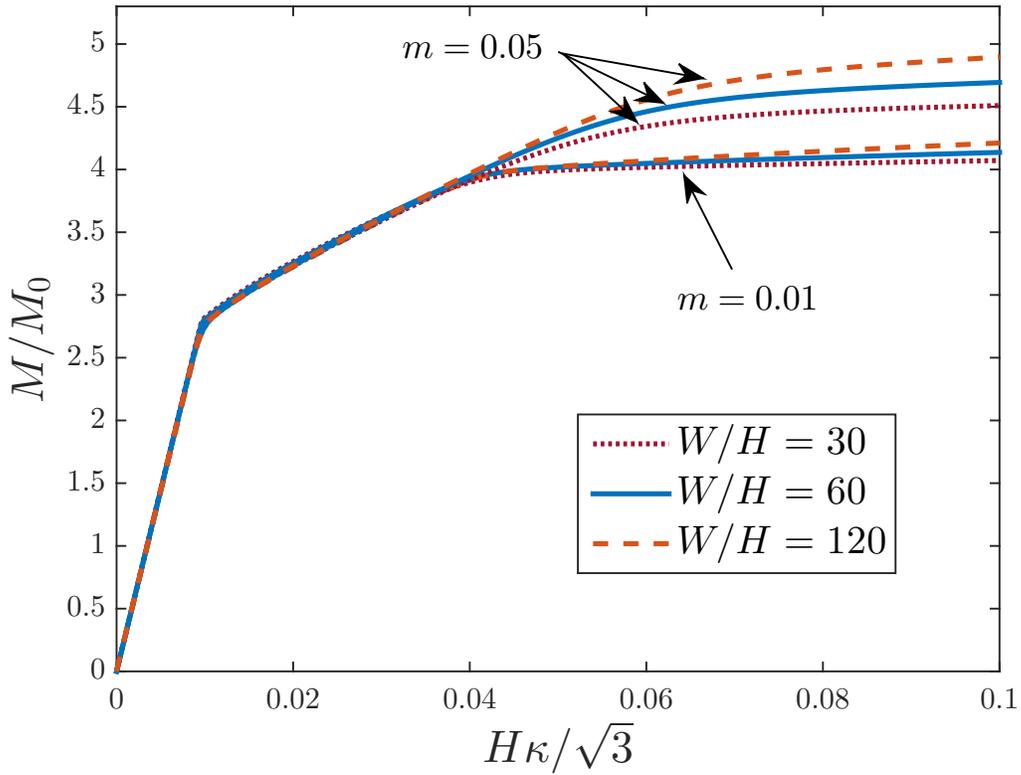}}
\caption{Normalized moment versus curvature for different foil lengths with $\chi=2/3$.}
\label{fig:FoilLength}
\end{figure}

The behavior so far described leads to a bending response dependent on the foil length $W$, for a given $H$.
This can be seen clearly in Fig. \ref{fig:FoilLength}, where the bending moment $M$ is plotted versus 
the applied curvature for $W/H=30$, $W/H=60$, and $W/H=120$. We consider two values of $m$ to investigate the response by gradually approaching rate-independence. Here and henceforth, $M$ is normalized by $M_0 = \sigma_0 H^2/(6 \sqrt{1-\nu + \nu^2})$, defining initial yielding in conventional rate-independent, 
von Mises plasticity.

\begin{figure}[H]
\centering
\noindent\makebox[\textwidth]{%
\includegraphics[scale=0.95]{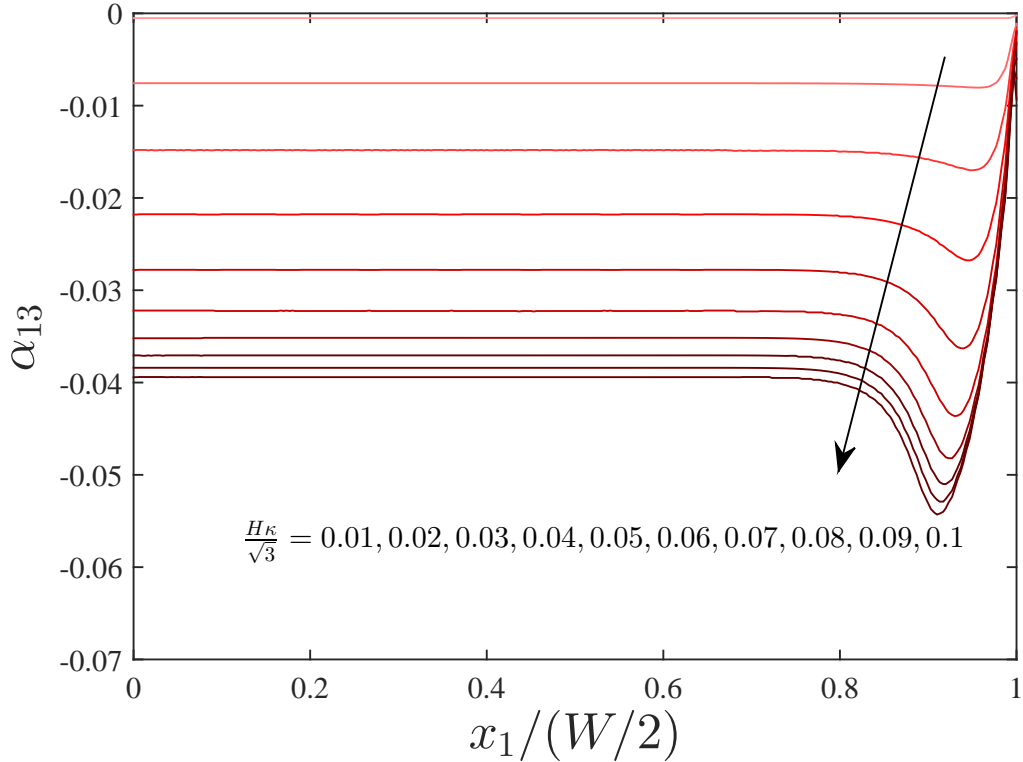}}
\caption{Variation of $\alpha_{13}$ along $x_1$ (at $x_2=H/4$) for $W/H=30$ at different applied curvature values. The following material properties are adopted: $\chi=2/3$ and $m=0.05$.}
\label{fig:A13}
\end{figure}

%As shown in Fig. \ref{fig:FoilLength}, 
The response is more compliant as $W$ diminishes, this behavior becoming irrelevant when the rate effects are small. After the initial elastic regime, delayed plasticity initiates at about $M/M_0\approx 2.8$ as a consequence of the dissipative gradient effects. A hardening regime follows due to the build-up of free energy associated with Nye's tensor until the response eventually saturates. The asymptotic values of $M$ are given by the minimizing field of functional (\ref{eq:MPI}) under the constraint $\dot{\boldsymbol{\alpha}}=\boldsymbol{0}$. As shown in Fig. \ref{fig:A13},
for large enough $\kappa$ Nye's tensor becomes insensitive to further increase of $\kappa$.
%as \hl{$\kappa$} increases $\alpha_{13}$ approaches a limit level that is insensitive to further increases of $\kappa$.

More insight can be gained by inspection of the first scalar equation included in the tensorial higher-order balance equation \eqref{ho1_iso}, whose leading terms are:
\beq
{\Sigma\over \dot E^p}\dot\varepsilon_{11}^p - \sigma_{11} +{1\over 2}\sigma_{33}
+\mu\ell^2(\underbrace{\gamma_{12,1}-\varepsilon_{11,2}^p}_{\alpha_{13}})_{,2} \approx 0
\label{discWH}
\eeq

\noindent where $\sigma_{33}$ basically depends on $\sigma_{11}$ through the hindered contraction along the $x_3$ direction, and we have neglected the terms $(\varepsilon_{22,1}^p-\gamma_{21,2})_{,1}$, $\varepsilon_{33,22}^p$, and $\varepsilon_{33,11}^p$. As already demonstrated, a quite large $\gamma_{12}$ must develop at the foil end to satisfy %, whenever there is plasticity, 
%condition \eqref{mfreethe12W}, that is 
condition $\alpha_{13}=0$. At a certain level of 
%applied curvature within the plastic range, 
$\kappa$,
it may become energetically convenient for the model to accommodate further increments 
$\dot{u}_1$ by developing almost only $\varepsilon_{11}^p$ in the foil end region, where $\gamma_{12}$ is already conspicuous and may further develop in such a way as to make
$\dot\alpha_{13}\approx 0$ pointwise in that domain. Thus, continued plastic deformation while preserving a constant Nye's tensor field leads to confinement of deformation close to the foil edge. In fact, examination of 
equations \eqref{nyeplanestrain}--\eqref{mfreeeps11H} reveals that a constant Nye's tensor field hinders a one-dimensional structure of the solution. % at the central part. 
Let us emphasize that, under the boundary conditions here concerned, this behavior is not observed in GP theories whose primal higher-order kinematic variables just consist of the plain gradient of $\bfgam$ 
(or $\bfeps^p$) and its rate. In fact, in such GP theories the bending solution is in terms of the direct plastic 
strain components only,
which turn out to be independent of $x_1$.
\par
\begin{figure}[H]
\centering
\noindent\makebox[\textwidth]{%
\includegraphics[scale=0.95]{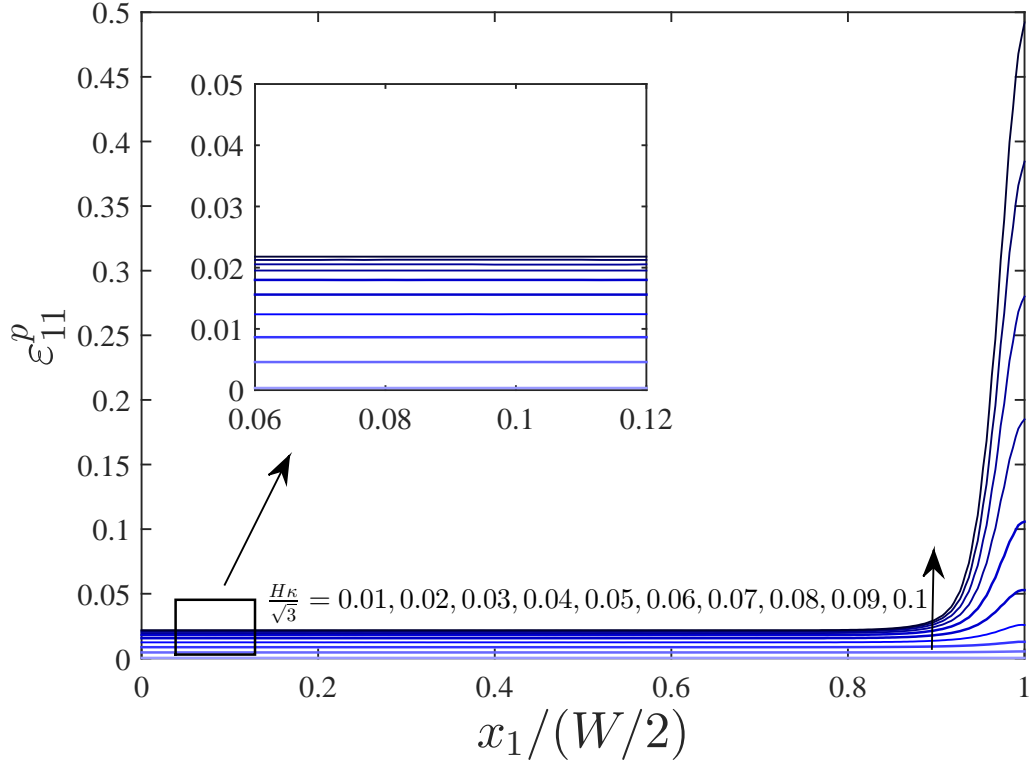}}
\caption{Variation of $\varepsilon_{11}^p$ along $x_1$ (at $x_2=H/4$) for $W/H=30$ at different applied curvature values. The following material properties are adopted: $\chi=2/3$ and $m=0.05$.}
\label{fig:Ep11B}
\end{figure}

Fig. \ref{fig:Ep11B} displays $\varepsilon_{11}^p(x_1,x_2=H/4)$ for various $\kappa$. It is observed that after a certain 
value of $\kappa$ is reached, further increasing it leads to concentration of $\varepsilon_{11}^p$ in the foil end region. %while little deformation occurs in the central part.

This behavior is particularly favorable in the rate-independent case without isotropic hardening ($N=0$) and
implies no further appreciable increase of 
longitudinal elastic strain, %and consequently of $\sigma_{11}$, 
in turn leading to vanishing increments of $M$.
Under these circumstances, {\it since the foil end regions where $\gamma_{12}$ significantly develops is proportional to the foil height $H$, not to 
the foil length $W$}, longer foils are subject to larger plastic flow at the foil end, for a given 
{\it applied curvature} $\kappa$. In fact, as evident from equation \eqref{conv_u1}, for a given $\kappa$ 
the {\it applied displacement} $u_1$ is proportional to $W$,
whereas, in the picture above, $u_1$ is then distributed in the field $\varepsilon_{11}^p$ solely over the foil end region.
 
Instead, if either rate-dependence or isotropic hardening are accounted for, 
$\Sigma$ in the first term of relation \eqref{discWH} 
increases with plasticity, so that, if the behavior above described is still energetically convenient, such that the fourth term remains of \eqref{discWH} small, there is the need of an increase of
%$\sigma_{11}$ (and, about proportionally, of $\sigma_{33}$) 
the Cauchy stress
to satisfy the higher-order balance equation. 
This makes $M$ larger and 
leads to the observed behavior that {\it shorter foils have softer mechanical response} in the viscoplastic (or isotropic hardening) case. In fact, because of the above described way to develop plasticity, shorter the foil, at a given 
$\kappa$, lower $\dot E^p$ due to a further increase in $\kappa$. Hence, for a shorter foil there is less hardening in the 
$M$ vs $\kappa$ response. Consequently, $\gamma_{12}$ in the foil end region increases with $W$ for a given 
$\kappa$ and the plastic spin
may play a major role in slender foils
(e.g., $W/H=120$ as in the experimental work of \citealp{SE98}). %This enhanced plasticity for longer beams must imply larger overall loads within the context of either isotropic hardening or, as in the present analysis, visco-plastic hardening.
Let us finally remark that this behavior is the result of the {\it unique} solution of the analyzed micro-bending problem, so that it is unrelated to any {\it localization} phenomenon. 
Also, we remain agnostic on whether this behavior describes what really
occurs at microfree boundaries subject to a direct plastic strain component, normal to the boundary, having a non-vanishing gradient 
along a tangential
direction. Hopefully, in the future, new experiments will shed light on this.
\footnote{
Unfortunately, further insight may not be gained by comparing our predictions with the crystal plasticity predictions of \citet{YVDGG04} and \citet{YVDG05}, as in these works the foil edge regions are constrained to remain linear elastic at any curvature level.
}

\subsection{Influence of the unconventional material parameters on the micro-bending response}

\begin{figure}[H]
\centering
\noindent\makebox[\textwidth]{%
\includegraphics[scale=0.95]{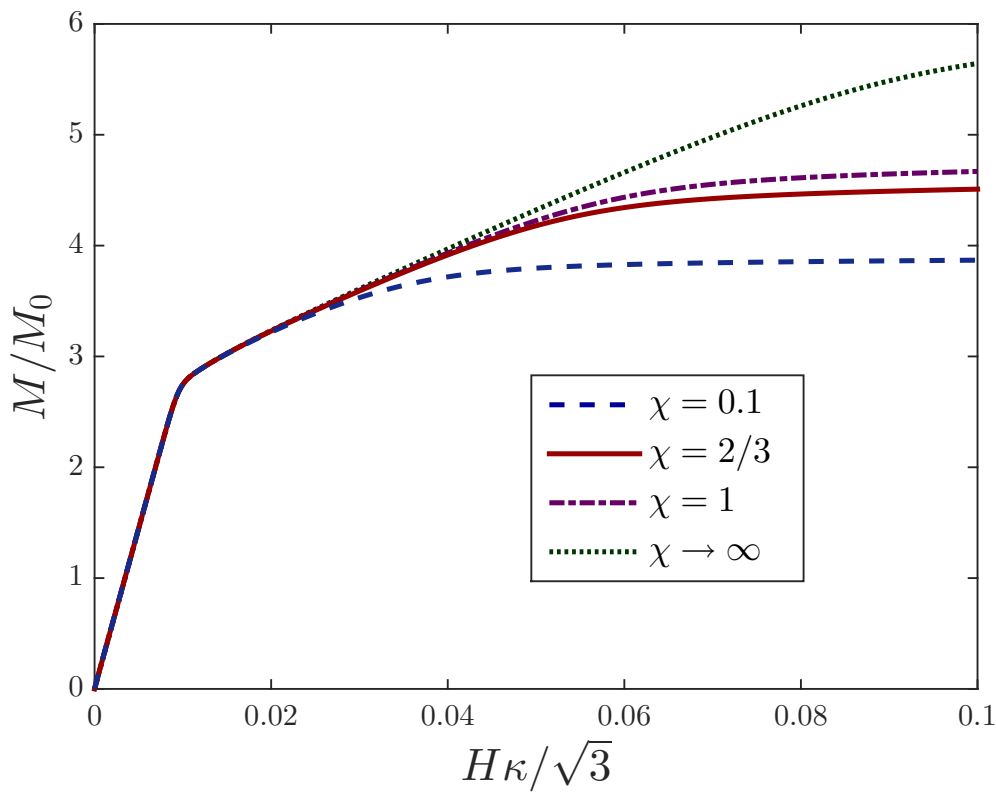}}
\caption{Normalized moment versus curvature for different values of $\chi$ with $m=0.05$.}
\label{fig:Plasticspin}
\end{figure}

The influence of $\chi$ 
in the mechanical response is examined for the reference ratio $W/H=30$ and results are reported in 
Fig. \ref{fig:Plasticspin}.

%As clearly observed in Fig. \ref{fig:Plasticspin}, 
It is observed that
increasing $\chi$ promotes hardening in later deformation stages.  More specifically, inspection of the higher-order balance equations (\ref{ho1_iso}) and (\ref{ho2_iso}) shows that augmenting $\chi$, while penalizing the plastic spin, leads to a larger defect stress, which plays the role of a backstress in equations (4) and (5) interpreted as a flow rule \citep{G04}. Hence, the increase in hardening with $\chi$ shown in Fig. 9 actually consists of an increase in the kinematic hardening related to GNDs.
\par
\begin{figure}[H]
\centering
\noindent\makebox[\textwidth]{%
\includegraphics[scale=0.95]{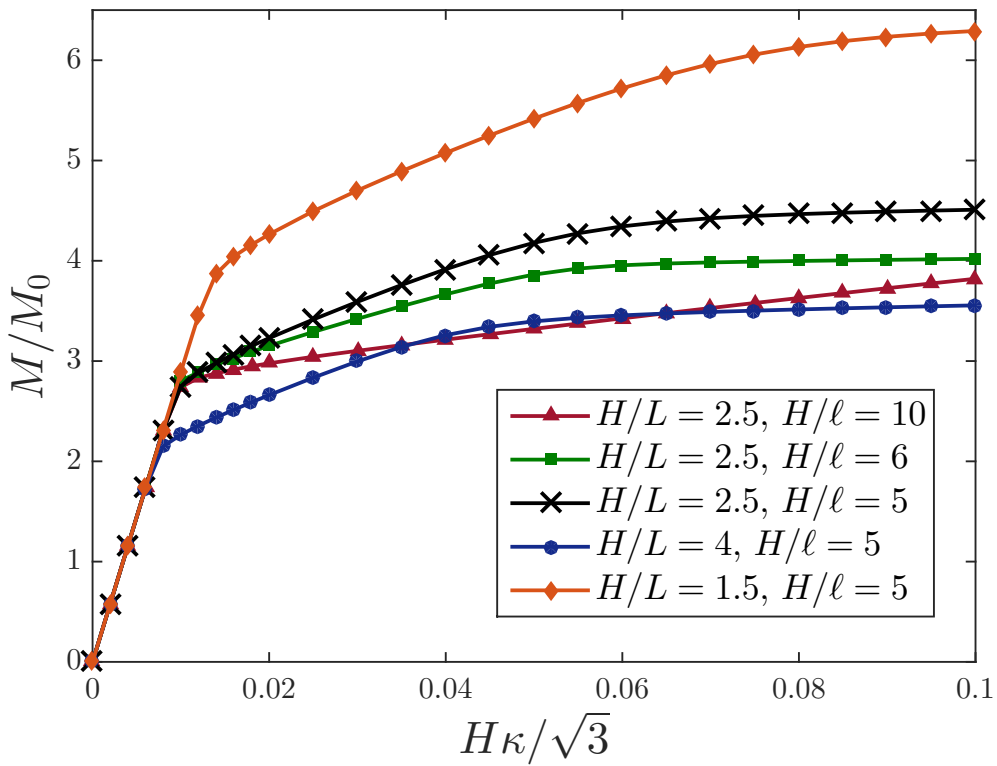}}
\caption{Normalized moment versus curvature for different values of $L$ and $\ell$. Other material parameters are: $\chi=2/3$ and $m=0.05$.}
\label{fig:DissipativeL}
\end{figure}

The role of the dissipative and energetic length scales in the $M$ vs $\kappa$ response has also been studied,
%and the results are reported 
as shown in Fig. \ref{fig:DissipativeL}.

%Fig. \ref{fig:DissipativeL} shows that, 
As expected,
the dissipative length scale $L$ governs the strengthening size effect: 
increasing $L$ leads to a clear rise in what is recognized as the ``initial yield moment''. It can also be appreciated %in Fig. \ref{fig:DissipativeL} 
that the energetic length scale $\ell$ governs the increase in the (kinematic) strain hardening with diminishing size. Therefore, the foregoing results show that,
by accounting for both energetic and dissipative higher-order contributions in \citet{G04} DGP theory, 
the present FE implementation 
can qualitatively reproduce the size effects observed in the experiments.\\

\subsection{Mechanical response under non-proportional loading}

{\it Non-incremental} dissipative higher-order terms (as referred to with the terminology used by \citealp{F14}) were introduced by \cite{G04} (see also \citealp{GU04, GA05}) in such a way as to ensure that stresses associated with unrecoverable plastic flow always result in positive plastic work, 
as stated by equation \eqref{sat_2nd} in the DGP here concerned. However, it has been very recently noticed \citep{F14,F15} that this may lead to a delay in plastic flow under certain non-proportional loading conditions, such a delay being referenced to as {\it elastic gap} by \citet{F14}.\\

The boundary value problem under study is characterized by imposing microhard boundary conditions at the foil top and bottom 
surfaces after a significant amount of plasticity has developed in bending 
under microfree boundary conditions. Such a switch of higher-order boundary conditions %, in the plastic regime, 
models the formation of passivation layers. A perfectly plastic foil of ratio $W/H=30$ is examined and the following material properties are adopted: $H/\ell=5$, $\chi=2/3$, $\sigma_0=200$ MPa, $\dot{\varepsilon}_0=0.02$ s$^{-1}$, $m=0.05$, $\nu=0.3$, and $\mu=26.3$ GPa.
In general, dislocations are forced to pile-up at the boundary by imposing the microhard boundary conditions \eqref{mh1} or \eqref{mh2}, depending on whether $L>0$ or $L=0$, respectively.
Here, we impose microhard conditions \eqref{mh1} because the case without dissipative higher-order effect, $H/L\to\infty$, is numerically treated by choosing an appropriately small positive value for $L$.
Finally, the microhard conditions \eqref{mh1}, in the plane strain case here of interest, turn out to imply 
\begin{equation}
\dot{\boldsymbol{\gamma}}=\bfzero \ \ \ \hbox{at}\ \ \ x_2 = \pm H/2
\end{equation}

\begin{figure}[H]
        \centering
        \begin{subfigure}[h]{0.49\textwidth}
                \centering
                \includegraphics[scale=0.45]{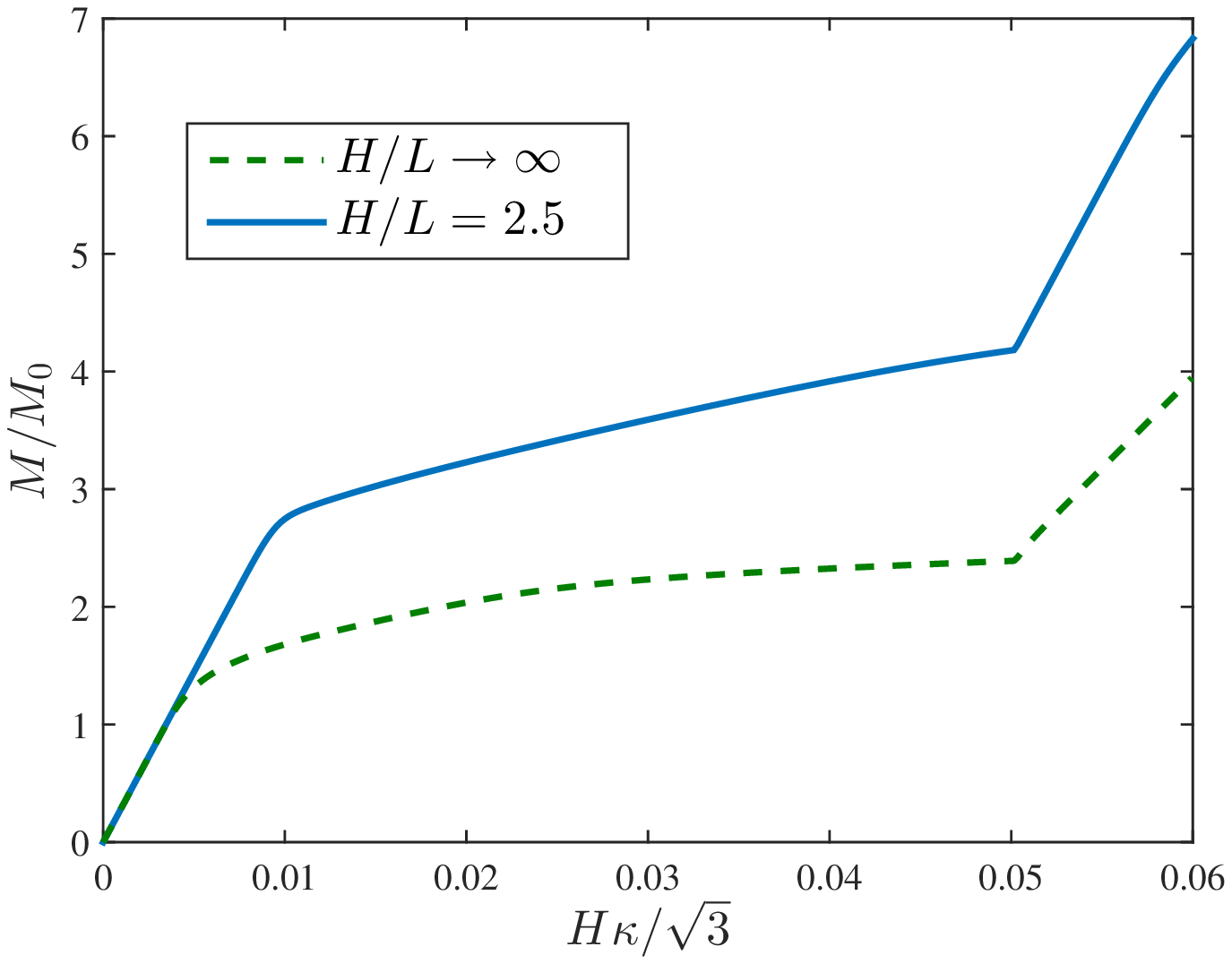}
                \caption{}
                \label{fig:Fig10}
        \end{subfigure}
        \begin{subfigure}[h]{0.49\textwidth}
                \centering
                \includegraphics[scale=0.45]{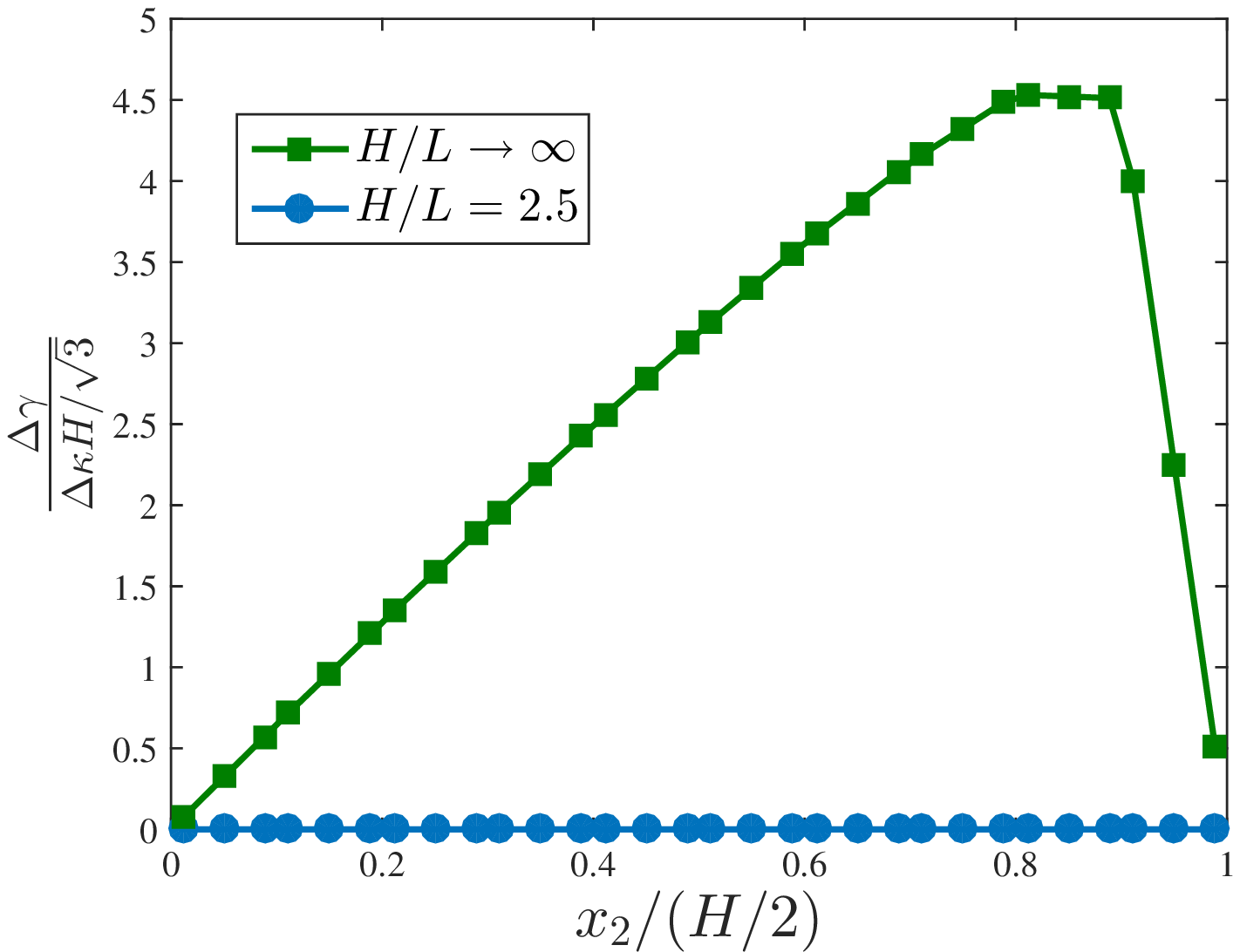}
                \caption{}
                \label{fig:Fig11}
        \end{subfigure}
       
        \caption{Effect of the application of a passivation layer: (a) Normalized moment versus curvature for different values of $L$ and (b) normalized plastic distortion increments along the thickness of the foil immediately after passivation. Other material parameters are $\chi=2/3$ and $m=0.05$.}\label{fig:FigElastic}
\end{figure}

Results obtained after switching the higher-order boundary conditions at $H \kappa / \sqrt{3} \approx 0.05$ are displayed in 
Fig. \ref{fig:FigElastic}, which clearly shows an abrupt stiffening at the formation of the passivation layers.

Qualitatively, the two options examined ($L \to 0$ or $L>0$) seem to lead to totally similar $M$ vs $\kappa$ responses. However, the two mechanical behaviors are very different, as observable in 
Fig. \ref{fig:Fig11}, showing
%the variation of an effective normalized measure of 
the incremental plastic distortion along the thickness of the mid-section ($x_1=0$) immediately after passivation. Here, the incremental plastic distortion is represented in terms of its modulus $\Delta \gamma = 
\Delta t |\dot{\boldsymbol{\gamma}}|$, with $\Delta t$ the time increment, 
and it is normalized by $\Delta \kappa 
= \Delta t\ \dot{\kappa}$.\\

Results reveal that setting $L>0$ leads to a purely elastic incremental response after 
formation of the passivation layer. This elastic gap after switching higher-order boundary conditions has been also numerically observed by \citet{BP15} in the torsion problem governed by DGP. %here concerned. 
As shown in Fig. \ref{fig:Fig11}, the elastic gap may be avoided by suppressing the 
unrecoverable higher-order term (i.e., by setting $L \to 0$). 
Our results provide further numerical evidence of the analytical findings of \citet{F14}. This may favor the ``incremental'' modeling approach suggested by \citet{H12}, where incremental relations between all the stress and strain variables are employed. Nevertheless, one should note that for $L \to 0$ the present formulation still has finite unrecoverable stresses constitutively conjugate to the plastic distortion rate, but not its gradient, which is the key issue pointed out by \citet{F14,F15}.\\

\section{Concluding remarks}
\label{Concluding remarks}

In small-scale plasticity,
the superior modeling capabilities associated with the constitutive inclusion of the plastic spin has 
recently 
encouraged significant interest in \textit{Distortion Gradient Plasticity} (DGP). 
In this work, we present a novel general purpose Finite Element (FE) framework for gradient theories involving the plastic spin, that is the skew-symmetric part of the plastic distortion. 
The proposed FE framework rests on two extremum principles and allows for an accurate modeling of both viscoplastic and rate-independent material responses. 
Such extremum principles extend to DGP those established by \citet{FW09b} for Strain Gradient Plasticity (SGP).\\

More specifically, we have focused on \citet{G04} DGP, which is characterized by the choice of Nye's dislocation density tensor
as primal higher-order kinematic variable, leading to a higher-order energetic stress, called defect stress, 
increasing with the plastic distortion incompatibility and governed by an energetic material length scale.

We have employed the novel FE framework for \citet{G04} DGP to implement general purpose plane strain elements.
The new numerical algorithm has been first %tly 
validated against literature results on the simple shear of a strip 
constrained between
bodies impenetrable to dislocations. 

Second, some specific features of \citet{G04} DGP theory have been analyzed by studying
the bending of thin metal foils. %, a paradigmatic
%benchmark in small-scale plasticity since the experimental results of \citet{SE98}.
Results show a strong influence of one shear component of the plastic distortion under microfree and conventional pure bending boundary conditions:
we have illustrated in detail the development of relevant plastic shear strain and spin required to 
compensate for the variation 
along the foil thickness of the longitudinal plastic strain.
This peculiarity is due to the 
form assumed by the microfree boundary conditions %in plane strain problems
in higher-order gradient plasticity based on Nye's tensor.

For a given foil thickness, this feature %, 
%given the imposed foil ends rotations in terms of longitudinal displacement, 
turns out in
a mechanical response exhibiting dependence on the foil length, with shorter foils being softer, 
if either rate-dependence or isotropic hardening are included in the modeling. 
This behavior
is also due to the imposed foil ends rotations, that are governed by 
the application of an average foil curvature.

The peculiar mechanism observed not only reveals a major role of the plastic spin but also indicates 
that analogous issues may be inherent to strain gradient {\it crystal} plasticity theories involving Nye's dislocation density 
tensor as primal higher-order kinematic variable.
In this context, counterintuitive coupling effects among slip systems have already been observed by \cite{BSPL13}.\\

The micro-bending benchmark has also been employed to investigate 
the existence of ``elastic gaps'' under non-proportional loading, as recently defined by \citet{F14}.
The proposed FE framework can predict that, in the present ``non-incremental'' DGP theory, a purely elastic incremental response follows passivation in the plastic regime. %, \hl{as expected from the analysis of \citet{F14}.}
%\hl{Here, passivation is the source of non-proportional loading as the passivated boundary is subject to switching from microfree to microhard conditions.} 
Critical experiments are needed to gain insight into the existence, or lack thereof, of the interruptions in plastic flow due to specific non-proportional loading conditions. Nevertheless, 
our FE analysis confirms that by assuming a vanishing value for the dissipative length scale governing the dissipative higher-order stress,
the present DGP formulation is free from such ``elastic gaps''. 
\par

\section{Acknowledgments}
\label{Acknowledge of funding}

Dr. Andrea Panteghini and Prof. Samuel Forest are acknowledged for helpful discussions. The authors gratefully acknowledge financial support from the Danish Council for Independent Research under the research career programme Sapere Aude in the project ``Higher Order Theories in Solid Mechanics". E. Mart\'{\i}nez-Pa\~neda also acknowledges financial support from the Ministry of Science and Innovation of Spain through grant MAT2011-28796-CO3-03, and the University of Oviedo through grant UNOV-13-PF and an excellence mobility grant within the International Campus of Excellence programme. L. Bardella additionally acknowledges financial support from the Italian Ministry of Education, University, and Research (MIUR).


\appendix
\section{Matrix operators for the discretization of the plastic variables}
\label{ShapeFunctions}

%\hl{
%$\boldsymbol{M}^{(i)}$ in equation \eqref{discgamma} is defined, in terms of the shape functions $N_i$, as:}
%\begin{equation}
%\boldsymbol{M}^{(i)}=\begin{bmatrix}
%  N_i & 0 & 0 & 0\\[0.3em]
%  0 & N_i & 0 & 0\\[0.3em]
%  0 & 0 & N_i & 0\\[0.3em]
%  0 & 0 & 0 & N_i\\[0.3em]
%  -N_i & -N_i & 0 & 0
% \end{bmatrix}
%\end{equation}
%\hl{Accordingly,}
The following matrices are defined in such a way as when they are multiplied by the column vector containing the four plastic distortion components of a node, say $[\gamma_{11}^{(i)},\ \gamma_{22}^{(i)},\ \gamma_{12}^{(i)},\ \gamma_{21}^{(i)}]^T$, they deliver its contribution to the vector fields containing the relevant components of 
the plastic strain $[\varepsilon_{11}^{p (i)},\ \varepsilon_{22}^{p (i)},\ \varepsilon_{12}^{p (i)},\ \varepsilon_{21}^{p (i)},
\ \varepsilon_{33}^{p (i)}]^T$, the plastic spin $[\vartheta_{12}^{p (i)},\ \vartheta_{21}^{p (i)}]^T$, the gradient of the plastic strain 

\noindent $[\varepsilon_{11,1}^{p (i)},\ \varepsilon_{11,2}^{p (i)},\ \varepsilon_{22,1}^{p (i)},
\ \varepsilon_{22,2}^{p (i)},\ \varepsilon_{12,1}^{p (i)},\ \varepsilon_{12,2}^{p (i)}$ $\varepsilon_{21,1}^{p (i)},\ 
\varepsilon_{21,2}^{p (i)},\ \varepsilon_{33,1}^{p (i)} ,\ \varepsilon_{33,2}^{p (i)}]^T$, and Nye's tensor $[\alpha_{13}^{(i)},\ \alpha_{23}^{(i)},\ \alpha_{31}^{(i)},\ \alpha_{32}^{(i)}]^T$, respectively:

\begin{equation}
\textnormal{sym} \, \boldsymbol{M}^{(i)}=\begin{bmatrix}
  N_i & 0 & 0 & 0\\[0.3em]
  0 & N_i & 0 & 0\\[0.3em]
  0 & 0 & \frac{1}{2} N_i & \frac{1}{2} N_i\\[0.3em]
  0 & 0 & \frac{1}{2} N_i & \frac{1}{2} N_i\\[0.3em]
  -N_i & -N_i & 0 & 0
 \end{bmatrix}
\end{equation}

\begin{equation}
\textnormal{skw} \, \boldsymbol{M}^{(i)}=\begin{bmatrix}
  0 & 0 & \frac{1}{2} N_i & -\frac{1}{2} N_i\\[0.3em]
  0 & 0 & -\frac{1}{2} N_i & \frac{1}{2} N_i
 \end{bmatrix}
\end{equation}

\begin{equation}
\textnormal{sym} \, \nabla \boldsymbol{M}^{(i)}=\begin{bmatrix}
  \frac{\partial N_i}{\partial x} & 0 & 0 & 0\\[0.3em]
  \frac{\partial N_i}{\partial y} & 0 & 0 & 0\\[0.3em]
   0 & \frac{\partial N_i}{\partial x} & 0 & 0\\[0.3em]
   0 & \frac{\partial N_i}{\partial y} & 0 & 0\\[0.3em]
   0 & 0 &\frac{1}{2} \frac{\partial N_i}{\partial x} & \frac{1}{2} \frac{\partial N_i}{\partial x}\\[0.3em]
   0 & 0 &\frac{1}{2} \frac{\partial N_i}{\partial y} & \frac{1}{2} \frac{\partial N_i}{\partial y}\\[0.3em]
   0 & 0 &\frac{1}{2} \frac{\partial N_i}{\partial x} & \frac{1}{2} \frac{\partial N_i}{\partial x}\\[0.3em]
   0 & 0 &\frac{1}{2} \frac{\partial N_i}{\partial y} & \frac{1}{2} \frac{\partial N_i}{\partial y}\\[0.3em]   
   - \frac{\partial N_i}{\partial x} & - \frac{\partial N_i}{\partial x} & 0 & 0 \\[0.3em]
     - \frac{\partial N_i}{\partial y} & - \frac{\partial N_i}{\partial y} & 0 & 0    
 \end{bmatrix}
\end{equation}

\begin{equation}
\textnormal{curl} \, \boldsymbol{M}^{(i)}=\begin{bmatrix}
  -\frac{\partial N_i}{\partial y} & 0 & \frac{\partial N_i}{\partial x} & 0\\[0.3em]
  0 & \frac{\partial N_i}{\partial x} & 0 & -\frac{\partial N_i}{\partial y}\\[0.3em]
   -\frac{\partial N_i}{\partial y} & -\frac{\partial N_i}{\partial y} & 0 & 0\\[0.3em]
   \frac{\partial N_i}{\partial x} & -\frac{\partial N_i}{\partial x} & 0 & 0
 \end{bmatrix}
\end{equation}


\begin{thebibliography}{00}


\bibitem[Ashby(1970)]{Ash70}
 Ashby, M.F.,
 1970.
 The deformation of plastically non-homogeneous materials.
 Philos. Mag.
 21,
 399-424.

\bibitem[Arsenlis and Parks(1999)]{AP99}
 Arsenlis, A., Parks, D.M.,
 2009.
 Crystallographic aspects of geometrically-necessary and statistically-stored dislocation density.
 Acta Mater.
 47,
 1597-1611.

\bibitem[Bardella(2009)]{B09}
 Bardella, L.,
 2009.
 A comparison between crystal and isotropic strain gradient plasticity theories with accent on the role of the plastic spin.
 Eur. J. Mech. A. Solids
 28,
 638-646.

\bibitem[Bardella(2010)]{B10}
 Bardella, L.,
 2010.
 Size effects in phenomenological strain gradient plasticity constitutively involving the plastic spin.
 Int. J. Eng. Sci.
 48,
 550-568.

\bibitem[Bardella and Giacomini(2008)]{BG08}
 Bardella, L., Giacomini, A.,
 2008.
 Influence of material parameters and crystallography on the size effects describable by means of strain gradient plasticity.
 J. Mech. Phys. Solids
 56,
 2906-2934.

\bibitem[Bardella and Panteghini(2015)]{BP15}
 Bardella, L., Panteghini, A.,
 2015.
 Modelling the torsion of thin metal wires by distortion gradient plasticity.
 J. Mech. Phys. Solids
 78,
 467-492.

\bibitem[Bardella et al.(2013)]{BSPL13}
 Bardella, L., Segurado, J., Panteghini, A., Llorca, J.,
 2013.
 Latent hardening size effect in small-scale plasticity.
 Modell. Simul. Mater. Sci. Eng.
 21,
 055009.

\bibitem[Bittencourt et al.(2003)]{B03}
 Bittencourt, E., Needleman, A., Gurtin, M.E., Van der Giessen, E.,
 2003.
 A comparison of nonlocal continuum and discrete dislocation plasticity predictions.
 J. Mech. Phys. Solids
 51,
 281-310.

\bibitem[Borg et al.(2006)]{BNFT06}
 Borg, U., Niordson, C.F., Fleck, N.A., Tvergaard, V., 
 2006.
 A viscoplastic strain gradient analysis of materials with voids or inclusions.
 Int. J. Solids Struct.
 43, 
 4906-4916.

\bibitem[Engelen et al.(2006)]{EFPG06}
 Engelen, R.A.B., Fleck, N.A., Peerlings, R.H.J., Geers, M.G.D.,
 2006.
 An evaluation of higher-order plasticity theories for predicting size effects and localisation.
 Int. J. Solids Struct.
 43,
 1857-1877.

\bibitem[Evans and Hutchinson(2009)]{EH09}
 Evans, A.G., Hutchinson, J.W.,
 2009.
 A critical assessment of theories of strain gradient plasticity.
 Acta Mater.
 57,
 1675-1688.

\bibitem[Fleck and Hutchinson(1997)]{FH97}
 Fleck, N.A., Hutchinson, J.W.,
 1997.
 Strain gradient plasticity.
 Adv. Appl. Mech.
 33,
 295-361.

\bibitem[Fleck and Hutchinson(2001)]{FH01}
 Fleck, N.A., Hutchinson, J.W.,
 2001.
 A reformulation of strain gradient plasticity.
 J. Mech. Phys. Solids
 41,
 1825-1857.

\bibitem[Fleck et al.(2014)]{F14}
 Fleck, N.A., Hutchinson, J.W., Willis, J.R.,
 2014.
 Strain gradient plasticity under non-proportional loading.
 Proc. R. Soc. London, Ser. A	
 470,
 20140267.
 
\bibitem[Fleck et al.(2015)]{F15}
 Fleck, N.A., Hutchinson, J.W., Willis, J.R.,
 2015.
 Guidelines for Constructing Strain Gradient Plasticity Theories.
 J. Appl. Mech.
 82,
 071002 1-10. 

\bibitem[Fleck and Willis(2009a)]{FW09a}
 Fleck, N.A., Willis, J.R.,
 2009a.
 A mathematical basis for strain-gradient plasticity theory - part I: Scalar plastic multiplier.
 J. Mech. Phys. Solids
 57,
 161-177.

\bibitem[Fleck and Willis(2009b)]{FW09b}
 Fleck, N.A., Willis, J.R.,
 2009b.
 A mathematical basis for strain-gradient plasticity theory - part II: tensorial plastic multiplier.
 J. Mech. Phys. Solids
 57,
 1045-1057.

\bibitem[Fleck et al.(1994)]{F94}
 Fleck, N.A., Muller, G.M., Ashby, M.F., Hutchinson, J.W.,
 1994.
 Strain gradient plasticity: theory and experiment.
 Acta Metall. Mater.
 42,
 475-487.
 
\bibitem[Forest and Gu\'{e}ninchault(2013)]{FG13}
 Forest, S., Gu\'{e}ninchault, N.,
 2013.
 Inspection of free energy functions in gradient crystal plasticity.
 Acta Mech. Sin.
 29,
 763-772. 
 
\bibitem[Fredriksson and Gudmundson(2005)]{FG05}
 Fredriksson, P., Gudmundson, P., 
 2005.
 Size-dependent yield strength of thin films.
 Int. J. Plast.
 21, 
 1834-1854.


\bibitem[Gao et al.(1999)]{G99}
 Gao, H., Huang, Y., Nix, W.D., Hutchinson, J.W.,
 1999.
 Mechanism-based strain gradient plasticity – I. Theory.
 J. Mech. Phys. Solids
 47,
 1239-1263.
 
\bibitem[Garroni et al.(2010)]{G10}
 Garroni, A., Leoni, G., Ponsiglione, M.,
 2010.
 Gradient theory for plasticity via homogenization of discrete dislocations.
 J. Eur. Math. Soc.
 12,
 1231–1266. 

\bibitem[Gudmundson(2004)]{GU04}
 Gudmundson, P.,
 2004.
 A unified treatment of strain gradient plasticity.
 J. Mech. Phys. Solids
 52,
 1379-1406.

\bibitem[Gurtin(2004)]{G04}
 Gurtin, M.E.,
 2004.
 A gradient theory of small-deformation isotropic plasticity that accounts for the Burgers vector and for dissipation due to plastic spin.
 J. Mech. Phys. Solids
 52,
 2545-2568.

\bibitem[Gurtin and Anand(2005)]{GA05}
 Gurtin, M.E., Anand, L.,
 2005.
 A theory of strain-gradient plasticity for isotropic, plastically irrotational materials. Part I: small deformations.
 J. Mech. Phys. Solids
 53,
 1624-1649.

\bibitem[Gurtin and Anand(2009)]{GA09}
 Gurtin, M.E., Anand, L.,
 2009.
 Thermodynamics applied to gradient theories involving the accumulated plastic strain: The theories of Aifantis and Fleck \& Hutchinson and their generalization.
 J. Mech. Phys. Solids
 57,
 405-421.
 
\bibitem[Gurtin and Needleman(2005)]{GN05}
 Gurtin, M.E., Needleman, A.,
 2005.
 Boundary conditions in small-deformation, single-crystal plasticity that account for the burgers vector.
 J. Mech. Phys. Solids
 53,
 1-31. 

\bibitem[Hutchinson(2012)]{H12}
 Hutchinson, J.W.,
 1994.
 Generalizing $J_2$ flow theory: fundamental issues in strain gradient plasticity.
 Acta Mech. Sin.
 28,
 1078-1086.

\bibitem[Idiart et al.(2009)]{I09}
 Idiart, M.I., Deshpande, V.S., Fleck, N.A., Willis, J.R.,
 2009.
 Size effects in the bending of thin films.
 Int. J. Eng. Sci.
 47,
 1251-1264.

\bibitem[Klusemann et al.(2013)]{K13}
 Klusemann, B., Svendsen, B., Vehoff, H.,
 2013.
 Modeling and simulation of deformation behavior, orientation gradient development and heterogeneous hardening in thin sheets with coarse texture.
 Int. J. Plast.
 50,
 109-126.
 
\bibitem[Lancioni et al.(2015)]{L15}
 Lancioni, G., Yal\c{c}inkaya, T., Cocks, A.,
 2015.
 Energy-based non-local plasticity models for deformation patterning, localization and fracture.
 Proc. R. Soc. London, Ser. A	
 471,
 20150275.

\bibitem[Legarth and Niordson(2010)]{LN10}
 Legarth, B.N., Niordson, C.F.,
 2010.
 Debonding failure and size effects in micro reinforced composites.
 Int. J. Plast.
 26,
 149-165.

\bibitem[Liu et al.(2005)]{L05}
 Liu, B., Huang, Y., Li, M., Hwang, K.C., Liu, C.,
 2005.
 A study of the void size effect based on the Taylor dislocation model.
 Int. J. Plast.
 21,
 2107-2122.

\bibitem[Mart\'{\i}nez-Pa\~neda and Beteg\'{o}n(2015)]{MB15}
 Mart\'{\i}nez-Pa\~neda, E., Beteg\'{o}n, C.,
 2015.
 Modeling damage and fracture within strain gradient plasticity.
 Int. J. Solids Struct.
 59,
 208-215.
 
\bibitem[Mart\'{\i}nez-Pa\~neda and Niordson(2016)]{MN16}
 Mart\'{\i}nez-Pa\~neda, E., Niordson, C.F.,
 2016.
 On fracture in finite strain gradient plasticity.
 Int. J. Plast.
 80,
 154-167. 

\bibitem[Moreau et al.(2005)]{Moreau05}
 Moreau, P., Raulic, M., P'ng, M.Y., Gannaway, G., Anderson, P., Gillin, W.P., Bushby, A.J., Dunstan, D.J.,
 2005.
 Measurement of the size effect in the yield strength of nickel foils.
 Philos. Mag. Lett.
 85,
 339-343.

\bibitem[Nielsen and Niordson(2014)]{NN14}
 Nielsen, K.L., Niordson, C.F.,
 2014.
 A numerical basis for strain-gradient plasticity theory: Rate-independent and rate-dependent formulations.
 J. Mech. Phys. Solids
 63,
 113-127.

\bibitem[Niordson(2003)]{N03}
 Niordson, C.F.,
 2003.
 Strain gradient plasticity effects in whisker-reinforced metals.
 J. Mech. Phys. Solids
 51,
 1863-1883.

\bibitem[Niordson(2007)]{N07}
 Niordson, C.F.,
 2007.
 Size-effects in porous metals.
 Modell. Simul. Mater. Sci. Eng.
 15,
 51-60.

\bibitem[Niordson and Hutchinson(2011)]{NN11}
 Niordson, C.F., Hutchinson, J.W.,
 2011.
 Basic strain gradient plasticity theories with application to constrained film deformation.
 J. Mech. Mater. Struct.
 6,
 395-416.

\bibitem[Niordson and Legarth(2010)]{NL10}
 Niordson, C.F., Legarth, B.N.,
 2010.
 Strain gradient effects on cyclic plasticity.
 J. Mech. Phys. Solids
 58,
 542-557.
 
\bibitem[Nix and Gao(1998)]{NG98}
 Nix, W.D., Gao, H.,
 1998.
 Indentation size effects in crystalline materials: a law for strain gradient plasticity.
 J. Mech. Phys. Solids
 46,
 411-425.   

\bibitem[Nye(1953)]{N53}
 Nye, J.F.,
 1953.
 Some geometrical relations in dislocated crystals.
 Acta Metall.
 1,
 153-162.
 
\bibitem[Ohno and Okomura(2007)]{OO07}
 Ohno, N., Okumura, D.,
 2007.
 Higher-order stress and grain size effects due to self-energy of geometrically necessary dislocations.
 J. Mech. Phys. Solids
 55,
 1879-1898. 

\bibitem[Ostien and Garikipati(2008)]{OG08}
 Ostien, J., Garikipati, K.,
 2008.
 Galerkin method for an incompatibility-based strain gradient plasticity theory,
 in: IUTAM Symposium on
Theoretical, Computational and Modelling Aspects of Inelastic Media, Springer, Netherlands,
 217-226.
 
\bibitem[Poh(2013)]{P13}
 Poh, L.H.,
 2013.
 Scale transition of a higher order plasticity model - A consistent homogenization theory from meso to macro.
 J. Mech. Phys. Solids
 61,
 2692-2710.
 
\bibitem[Poh and Peerlings(2016)]{PP16b}
 Poh, L.H., Peerlings, R.H.J.,
 2016.
 The plastic rotation effect in an isotropic gradient plasticity model for applications at the meso scale.
 Int. J. Solids Struct.
 78-79,
 57-69.
 
\bibitem[Poh and Phan(2016)]{PP16}
 Poh, L.H., Phan, V.T.,
 2016.
 Numerical implementation and validation of a consistently homogenized higher order plasticity model.
 Int. J. Numer. Methods Eng.
 106, 
 454-483.
 
\bibitem[Polizzotto(2011)]{Pol11}
 Polizzotto, C.,
 2011.
 Size effects on the plastic collapse limit load of thin foils in bending and thin wires in torsion.
 Eur. J. Mech. A. Solids
 30,
 854-864.

\bibitem[Qu et al.(2006)]{Q06}
 Qu, S., Huang, Y., Pharr, G.M., Hwang, K.C.,
 2006.
 The indentation size effect in the spherical indentation of iridium: a study via the conventional theory of mechanism-based strain gradient plasticity.
 Int. J. Plast.
 22,
 1265-1286.

\bibitem[St\"{o}lken and Evans(1998)]{SE98}
 St\"{o}lken, J.S., Evans, A.G.,
 1998.
 A microbend test method for measuring the plasticity length scale.
 Acta Mater.
 46,
 5109-5115.

\bibitem[Wieners and Wohlmuth(2011)]{WW11}
 Wieners, C., Wohlmuth, B.,
 2011.
 A primal-dual finite element approximation for a nonlocal model in plasticity.
 SIAM J. Numer. Anal.
 49,
 692-710.

\bibitem[Yefimov and Van der Giessen(2005)]{YVDG05}
 Yefimov, S., Van der Giessen, E.,
 2005.
 Multiple slip in a strain-gradient plasticity model motivated
 by a statistical-mechanics description of dislocations.
 Int. J. Solids Struct.
 42,
 3375-3394.

\bibitem[Yefimov et al.(2004)]{YVDGG04}
 Yefimov, S., Van der Giessen, E., Groma, I.,
 2004.
 Bending of a single crystal: discrete dislocation and
 nonlocal crystal plasticity simulations.
 Modelling Simul. Mater. Sci. Eng.
 12,
 1069-1086.
 
\end{thebibliography}


\end{document}